\documentclass[%
 reprint,
superscriptaddress,
 amsmath,amssymb,
 aps,
]{revtex4-2}

\usepackage{graphicx}
\usepackage{dcolumn}
\usepackage{bm}
\usepackage[utf8]{inputenc}
\usepackage[T1]{fontenc}
\usepackage{mathptmx}
\usepackage{etoolbox}
\usepackage{amsmath}
\usepackage{braket}
\usepackage{dsfont}
\usepackage{units}
\usepackage{placeins}
\usepackage{comment}
\usepackage{nicematrix}

\begin{document}

\preprint{APS/123-QED}

\title{Theoretical development of a new spin filter generation}

\author{N.~Faatz}
\email{n.faatz@fz-juelich.de}

\affiliation{
 GSI, Helmholtzzentrum für Schwerionenforschung, Darmstadt, Germany 
}%
\affiliation{
 III. Physikalisches Institut B, RWTH Aachen University, Aachen, Germany}%
\affiliation{
 Institut für Kernphysik, Forschungszentrum Jülich, Jülich, Germany
}%
\author{R.~Engels}%
\email{r.w.engels@fz-juelich.de}
\affiliation{
 GSI, Helmholtzzentrum für Schwerionenforschung, Darmstadt, Germany 
}%
\affiliation{
 Institut für Kernphysik, Forschungszentrum Jülich, Jülich, Germany
}%
\author{B.~Breitkreutz}
 
\affiliation{
 GSI, Helmholtzzentrum für Schwerionenforschung, Darmstadt, Germany 
}%
\author{H.~Soltner}
\affiliation{ Zentralinstitut für Engineering, Elektronik und Analytik, Forschungszentrum Jülich, Jülich, Germany}%
\author{C.~Kannis}
\affiliation{ Institut für Laser- und Plasmaphysik, Heinrich-Heine-Universität Düsseldorf, Düsseldorf, Germany}

\date{\today}

\begin{abstract}
Since the early days of quantum mechanics hydrogen, as the simplest of all atoms, has been studied or used to investigate new physics. In parallel, this knowledge leads to different applications, e.g. a spin filter to separate metastable hydrogen atoms in single hyperfine substates with electron spin $m_s=\nicefrac{1}{2}$. Subsequently, this work provides the necessary theory as well as experimental conditions to build a new generation of spin filter which permits the separation of all four individual metastable hydrogen hyperfine states as well as for its isotopes in a corresponding beam.
\end{abstract}

\maketitle


\section{Introduction}
Hydrogen is the element with the simplest structure, since the nucleus comprises a single proton orbited by one electron. It is the only atom for which the Schrödinger equation can be solved analytically. The hydrogen atom and its energy corrections are described by a well-established theory, which makes it suitable for experiments addressing the effects of polarization. In this context the term polarization is defined as the average spin orientation for an ensemble of particles. This orientation refers to the alignment of the spin magnetic moment to an external magnetic field. To measure the polarization of a hydrogen beam in the metastable $2S_{\nicefrac{1}{2}}$ state a Lamb-shift polarimeter (LSP)~\cite{Lamb_pol,Ralf} is a useful instrument and has been successfully used for many years at the polarized internal target of the ANKE experiment~\cite{COSY}. One important part of the underlying detection method is the so called spin filter~\cite{Los_Alamos}. Its purpose is to distinguish metastable atoms with different spin configurations from each other. For the hydrogen atom four spin combinations are possible, defined by the hyperfine structure. Whereas the already existing LSP is only able to filter the two $\alpha$ states with electron spin up $\left(m_s=\nicefrac{1}{2}\right)$, the theory presented in this paper triggered the development of a second-generation spin filter to overcome this limitation and to filter all four states separately. The two unreachable states for the current set-up are called $\beta$ states and are characterized by having the electron spin down $\left(m_s=-\nicefrac{1}{2}\right)$. Several experiments may take advantage of this upgrade, as their analysis depends on one or both of the $\beta$ states. One application is the  BoB experiment, which aims at analyzing the bound beta decay~\cite{BOB}. The focus in this experiment is to determine the helicity of the anti-electron neutrino by measuring the occupation number of hydrogen atoms in the forbidden $\beta_3$ state $\left(\ket{F=1, m_F=-1}\right)$ after the rare neutron decay $n \rightarrow H_{2S} + \bar{\nu}_e$.

Moreover, the evidence for parity violation in the case of the metastable hydrogen atom comes within reach as the detection of the $\beta$ states is key~\cite{Weak_interaction}. Without any external electric fields, transitions from the $\beta$ states into the $2P_{\nicefrac{1}{2}}$ set are only possible due to the weak interaction, which violates the parity conservation. This would lead to a measurement of the Weinberg angles at very low energies.
\newline
Another experiment to make use of a beam of hydrogen atoms in the $\beta_4$ substate can be the search for axions~\cite{Axions}. While the classical spin filter can be used for the search of anthropic QCD dark matter axions at $10^{-7}$~eV that can induce a $\alpha_2 \rightarrow \alpha_1$ transition, this new spin filter might allow to observe transitions from the $\beta_4$ into $\alpha_1$ due to dark matter axions at energy levels of about $10^{-6}$~eV.

\begin{figure}[h!]
\scalebox{0.45}{\includegraphics{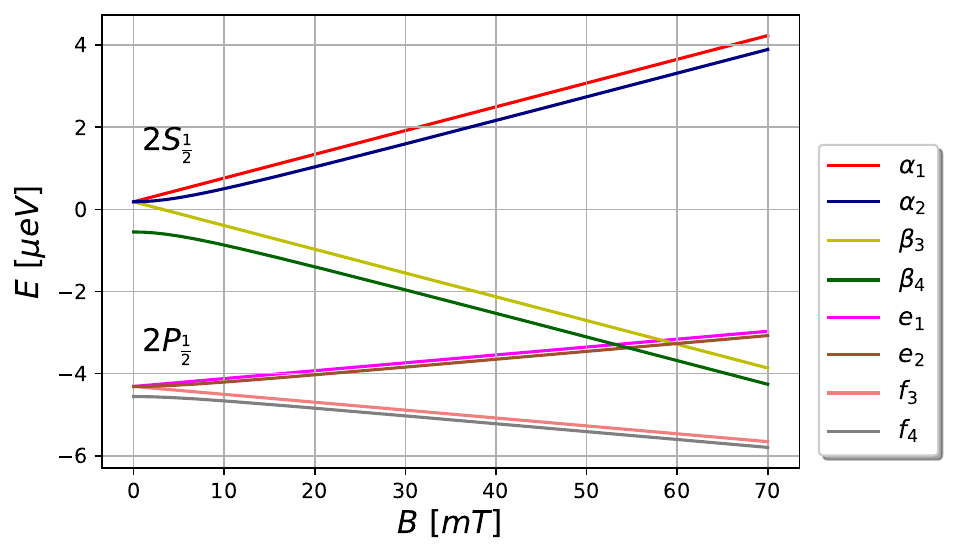}}
\caption{\label{BR_old} This plot shows the binding energies of the hyperfine substates (Breit-Rabi diagram) as function of an external homogeneous magnetic field for the metastable hydrogen $2S_{\nicefrac{1}{2}}$ and the $2P_{\nicefrac{1}{2}}$ set, respectively. The zero point on the $y$ axis is defined by the energy difference of both sets to each other. Including the fine splitting their binding energies should be equal, but the Lamb-shift~\cite{Lamb_shift}, as a product of QED corrections, separates these energy levels.}
\end{figure}
\FloatBarrier

\section{First-generation spin filter}
\begin{figure}[h!]
\scalebox{0.45}{\includegraphics{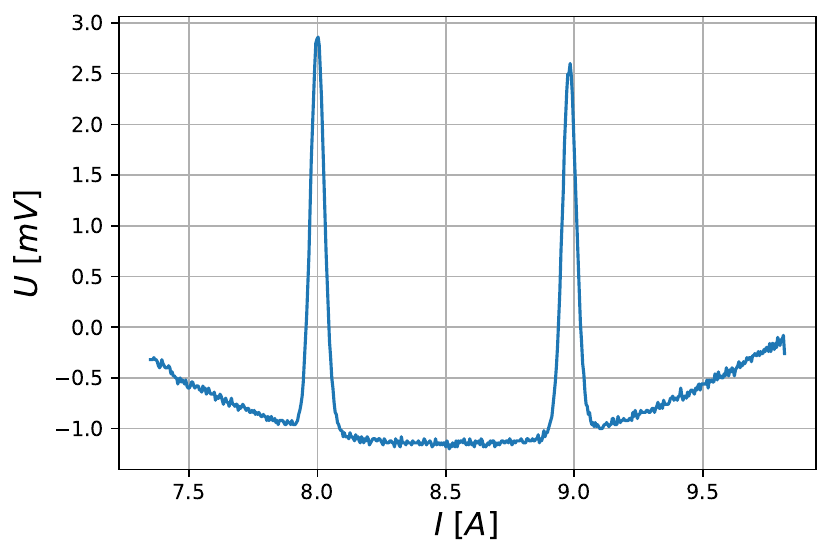}}
\caption{\label{Unpol_messung} The signal of a photomultiplier as function of the current inside the spin filter for an initial unpolarized hydrogen beam with a beam energy of $E_{beam}=1.5$ keV.}
\end{figure}

\begin{figure}[h!]
\scalebox{0.45}{\includegraphics{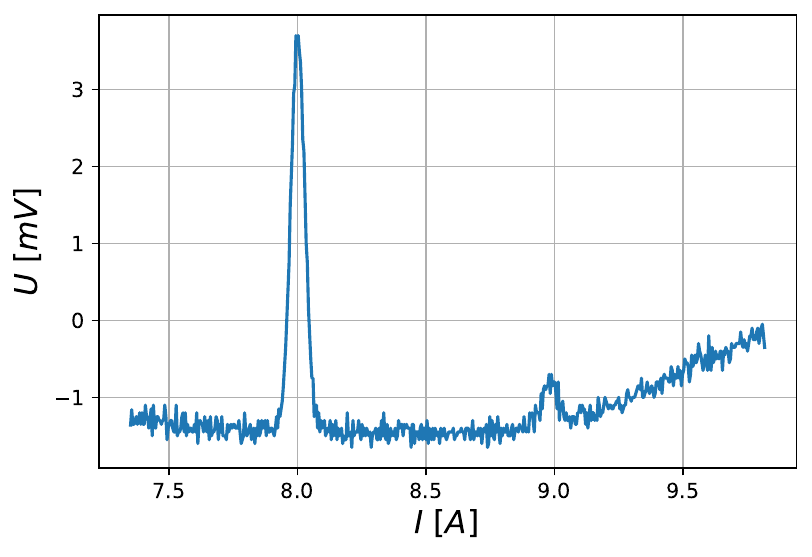}}
\caption{\label{a1_messung} The signal of a photomultiplier as function of the current inside the spin filter for an initial polarized hydrogen beam in the state $\alpha_1$ with a beam energy of $E_{beam}=1.5$ keV. From this spectrum the nuclear polarization of the incoming proton beam was determined as $P\approx0.72$}
\end{figure}

\begin{figure*}
\scalebox{0.7}{\includegraphics{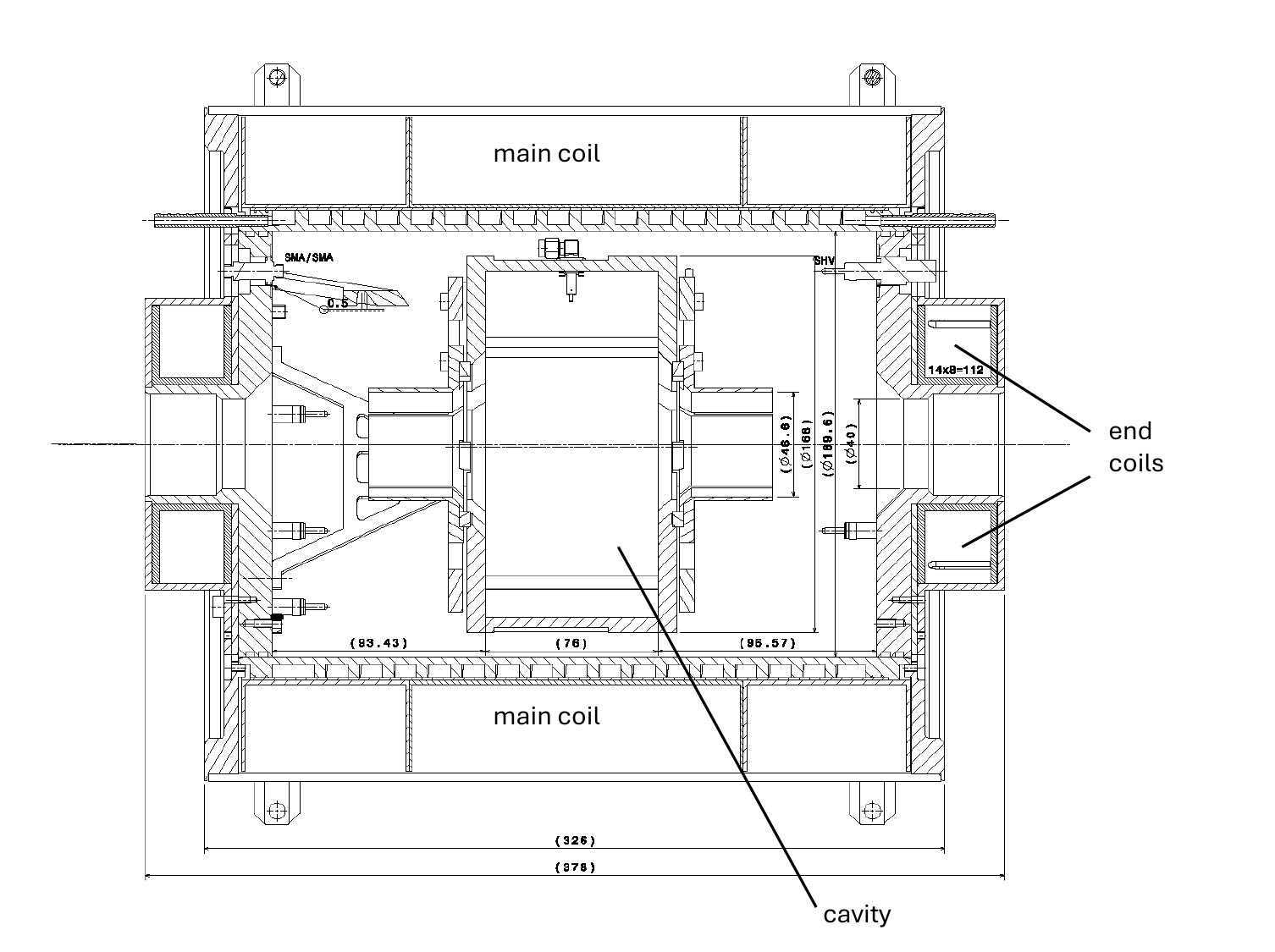}}
\caption{\label{Sketch} This two dimensional cross section shows the realization of the old type of spin filters. In the middle of the device sits the cavity. On the outside the magnetic field coils are placed.}
\end{figure*}

The already existing type of spin filter utilizes a static, homogeneous magnetic field along the beam direction. In addition, it features a cavity inside the magnetic field coils that provides a static electric field perpendicular to the beam axis as well as a resonant radio frequency at $f=1.60975$ GHz in the mode $\text{TM}_{0,1,0}$~\cite{RF,Los_Alamos,SF}. All these components are visualized in Fig.~\ref{Sketch}. The main coils produce the homogeneous magnetic field $B_0$, which is necessary to achieve the energy splitting shown in Fig.~\ref{BR_old} for the metastable hydrogen atoms. In between the cavity is located, which is divided into four isolated quadrants to apply the static electric field as well as the radio frequency. The old spin filter then utilizes the energy crossing at around $B\approx 57$~mT between the short lived $2P_{\nicefrac{1}{2}}$ set and the metastable $2S_{\nicefrac{1}{2}}$ set to induce electric dipole transitions via the static electric field also known as the Stark effect~\cite{Stark}. Definitions of the Breit-Rabi eigenstates are obtainable in the appendix~\ref{BR-States} as well as in table~\ref{tab_States}. This reduces the occupation numbers of the $\beta$ states, which are transferred to the $e$ states of the $2P_{\nicefrac{1}{2}}$ set. From there on the radio frequency can couple the $e$ states to the remaining $\alpha$ states. Only the corresponding $\alpha$ state above the crossing point then survives the time of flight. Thus, it is possible to control the lifetime of the single substates, i.e. to quench all into the ground state or to keep the occupation number of a single $\alpha$ state at dedicated magnetic fields around the crossing points (see Fig.~\ref{SF1}). Experimentally metastable hydrogen atoms are produced from a proton beam by charge exchange with cesium vapor, and their occupation numbers are then verified in the quenching chamber where the metastable hydrogen atoms are quenched into the ground state while the produced $Ly-\alpha$ photons are detected by a photomultiplier~\cite{Lamb_pol,Ralf}. Two of these measurements for a given spin filter situated at Forschungszentrum Jülich have been conducted and the results are given in Fig.~\ref{Unpol_messung} and Fig.~\ref{a1_messung}. In ideal measurements the peaks of an unpolarized beam should be of the same heights and the background a flat line. Therefore, this gives another motivation for improved simulations to compare these with measurements to identify the key parameters. Experiments showed that inhomogeneous magnetic fields produce unequal peak heights and reduce intensities while less radio frequency power leads to a more curved background. In addition, the strength of the static electric field influences the peak width, i.e. larger amplitudes corresponds to broader peaks.     
\FloatBarrier
\FloatBarrier
\begin{table}
\caption{\label{tab_States} The table defines the Breit-Rabi states at $n=2$ for hydrogen in the case of a vanishing external magnetic field $B_0=0$. In this case, they correspond to the hyperfine states expressed by the coupled representation $\ket{F,m_F}$.}
\begin{ruledtabular}
\begin{tabular}{ccccc}
 &$F$&$m_F$&$J$
 &$L$\\
\hline
$\alpha_1$& $1$ & $1$ & $\nicefrac{1}{2}$ & $0$\\
$\alpha_2$& $1$ & $0$ & $\nicefrac{1}{2}$ & $0$\\
$\beta_3$& $1$ & $-1$ & $\nicefrac{1}{2}$ & $0$\\
$\beta_4$& $0$ & $0$ & $\nicefrac{1}{2}$ & $0$\\
$e_1$& $1$ & $1$ & $\nicefrac{1}{2}$ & $1$\\
$e_2$& $1$ & $0$ & $\nicefrac{1}{2}$ & $1$\\
$f_3$& $1$ & $-1$ & $\nicefrac{1}{2}$ & $1$\\
$f_4$& $0$ & $0$ & $\nicefrac{1}{2}$ & $1$\\
$g_1$& $2$ & $2$ & $\nicefrac{3}{2}$ & $1$\\
$g_2$& $2$ & $1$ & $\nicefrac{3}{2}$ & $1$\\
$g_3$& $2$ & $0$ & $\nicefrac{3}{2}$ & $1$\\
$g_4$& $2$ & $-1$ & $\nicefrac{3}{2}$ & $1$\\
$g_5$& $2$ & $-2$ & $\nicefrac{3}{2}$ & $1$\\
$h_6$& $1$ & $1$ & $\nicefrac{3}{2}$ & $1$\\
$h_7$& $1$ & $0$ & $\nicefrac{3}{2}$ & $1$\\
$h_8$& $1$ & $-1$ & $\nicefrac{3}{2}$ & $1$\\
\end{tabular}
\end{ruledtabular}
\end{table}

\subsection{The framework}

The hydrogen atom, including the fine structure as well as the hyperfine interaction, is well described by the total angular momentum $\vec{F}=\mathds{1}\otimes\vec{J}+\vec{I}\otimes\mathds{1}$, with $\vec{I}$ being the nuclear spin and $\vec{J}$ the total angular momentum of the electron. In the presence of an external magnetic field $\vec{B}=B_0\hat{e}_z$ the eigenstates $\ket{m_F,F}$ evolve to the Breit-Rabi states~\cite{Breit-Rabi2}. These satisfy the eigenproblem of the following Hamiltonian describing hyperfine structure splitting and the external interaction of a magnetic field
\begin{equation}\label{BR-Ham}
H=A\frac{\vec{I}\cdot\vec{J}}{\hbar^2}+\left(g_J\mu_B\frac{\vec{J}}{\hbar}-g_I\mu_k\frac{\vec{I}}{\hbar}\right)\cdot\vec{B},
\end{equation}
with $\mu_{B,k}$ being the Bohr and nuclear magneton, respectively. The Landé g-factor is represented by $g_J$, and $g_I$ corresponds to the nuclear g-factor. Finally, $A$ represents the hyperfine splitting constant, which depends on the quantum numbers $n$, $L$ and $J$~\cite{QM2}. The solution to the eigenproblem in addition to the fine splitting $\left(FS\right)$ and the Lamb shift $\left(\Delta E_{Lamb}\right)$ for the key states in this paper can be found in appendix~\ref{BR-States} and is illustrated in Fig.~\ref{BR} as well as in Fig.~\ref{BR_old}. An important point is that in Fig.~\ref{BR} the magnetic field scale is so large, that only half of the eigenenergies can be resolved. The other half of energy levels with different nuclear spin are all very close to one of the visible levels. In the next step, the first-order Stark effect is applied to the Breit-Rabi eigenstates~\cite{Stark}. This eigenproblem needs to be addressed numerically for an electric field given by $\vec{E}=\varepsilon_0\hat{e}_x$. For small fields $\left(\varepsilon_0\leq 10^3\,\frac{\text{V}}{\text{m}}\right)$ the perturbation onto the energy levels is nearly negligible. In contrast to this, the electric dipole interaction couples states with $\Delta L=\pm 1$, i.e. the short-living $2P$ to the long-living $2S$ states, such that the evolution of the lifetime is of great importance. This lifetime can be assumed as an additional damping term $V_{Life}=-i\frac{\hbar}{2\tau}$~\cite{Los_Alamos,QM2}, which is one of the key principles used in the spin filter.
\section{The second-generation spin filter}
The second-generation spin filter is based on the same concept as the old one, but the behavior of the $\alpha$ and $\beta$ states is inverted, such that the $\alpha$ states perform transitions and the $\beta$ states are repopulated by an electromagnetic wave with the right amount of energy equivalent to the energy gap. To find energy crossings for the $\alpha$ states, the $2P_{\nicefrac{3}{2}}$ set needs to be taken into account. States $h_7$ and $g_4$ fulfil the necessary conditions, and the corresponding crossings are situated at magnetic fields around $B_0\approx429$ mT and the necessary radio frequency to couple to the $\beta$ states corresponds to $f_2=11.94059$ GHz.
\newline
\newline
To integrate both concepts into a single device, it is essential to consider key characteristics of the existing spin filter, i.e. the energy crossings between the $\beta$ and $e$ states at $54$ mT and $60.5$ mT. Therefore, the energy difference corresponds to a frequency of $f_1=1.60975$~GHz~\cite{Ralf}. Subsequently, by utilizing Eq.~\eqref{RF_equation} a radius of $R_1=71.3$~mm corresponds to the frequency $f_1$ for a cylindrically shaped cavity~\cite{RF}. In comparison, the crossings between the $\alpha_1$ and $h_7$, correspondingly $\alpha_2$ to $g_4$, are located at magnetic fields of $423$~mT and $429$~mT, respectively. This leads to a frequency of $f_2=11.94059$~GHz to fill the energy gap between the $\beta$ states and the crossing points. For the same radius, the $6$th harmonic $\text{TM}_{0,6,0}$ provides a frequency of $f_{\text{TM}_{0,6,0}}=12.093$~GHz requiring only minor tuning to align with the correct frequency. The corresponding radius to obtain the frequency $f_2=11.94059$~GHz is given by $R_2=72.2$~mm, resulting in a difference of only $\Delta R\approx0.9$~mm. This means that with minor changes the new frequency $f_2$ can be incorporated into the first-generation spin filter. 

\subsection{Theory}
\begin{figure}[h!]
\scalebox{0.45}{\includegraphics{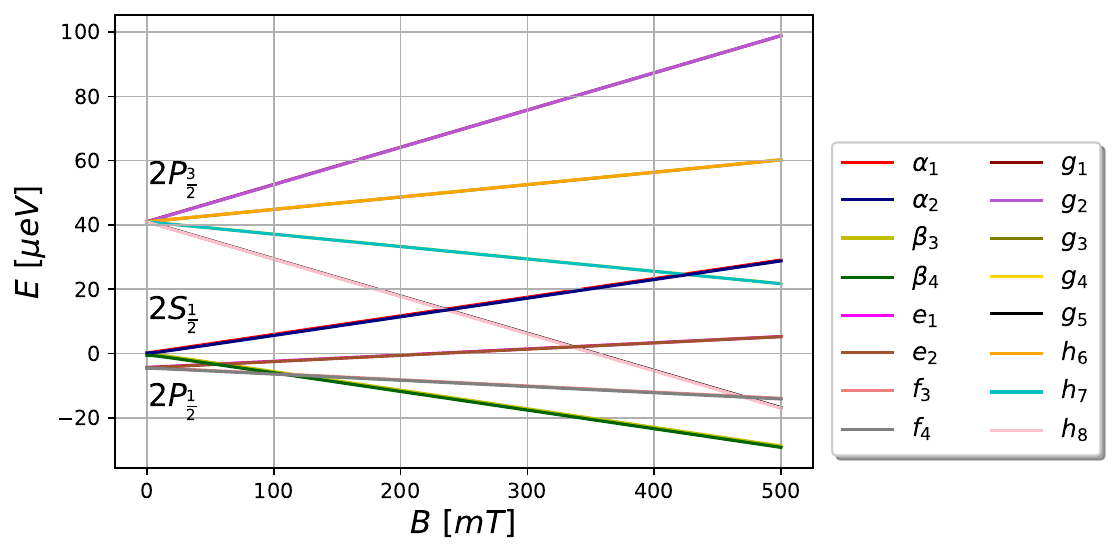}}
\caption{\label{BR} Eigenenergies of the Hamiltonian from Eq.~\eqref{BR-Ham} as function of the magnetic field amplitude for a static constant magnetic field in beam direction.}
\end{figure}

\begin{figure}[h!]
\scalebox{0.45}{\includegraphics{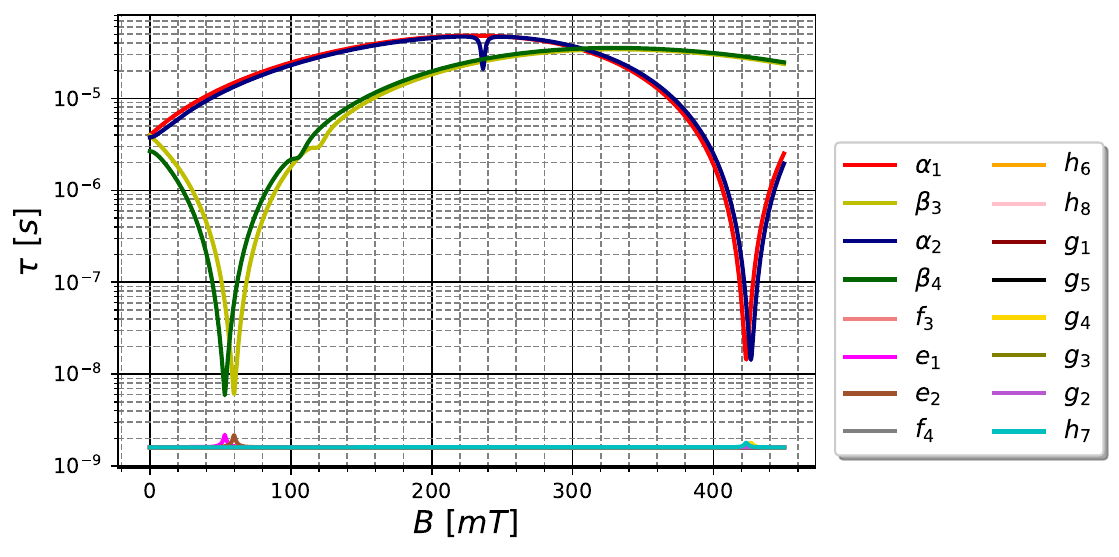}}
\caption{\label{Lifetimeplot} The lifetimes for the sixteen Breit-Rabi states are plotted as function of the magnetic flux density for a static constant magnetic field $B$ in beam direction in the presence of a static electric field perpendicular to the beam axis with a field strength of $\varepsilon_0=10^3\,\frac{\text{V}}{\text{m}}$.}
\end{figure}

\begin{figure}[h!]
\scalebox{0.45}{\includegraphics{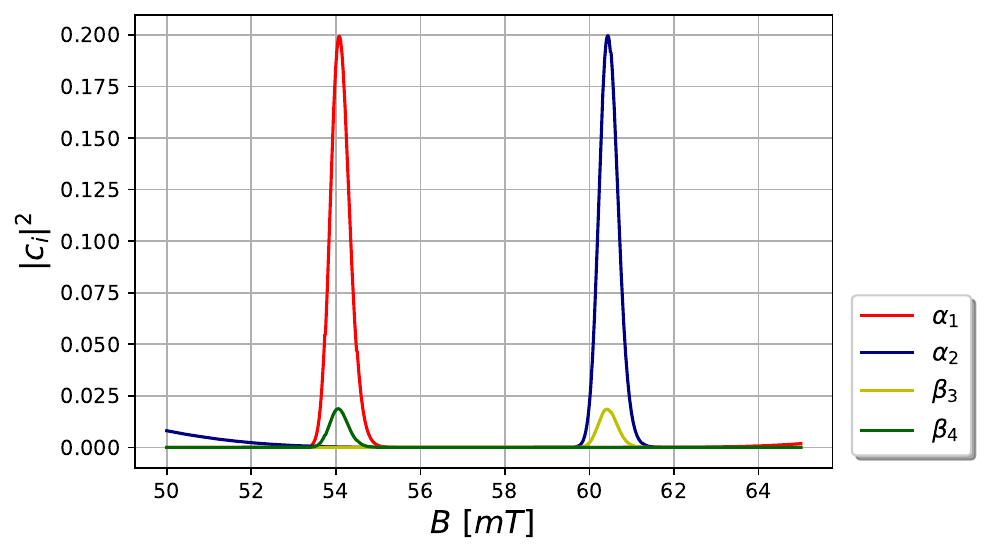}}
\caption{\label{SF1} The probability to find the single states are plotted as function of the magnetic field amplitude. In addition the radio frequency $\text{TM}_{0,1,0}$ mode has been used with a field strength of $\varepsilon_0=1.1\cdot10^3\,\frac{\text{V}}{\text{m}}$ and a static electric flux density of $\varepsilon_x=1.8\cdot10^3\,\frac{\text{V}}{\text{m}}$. Only the four relevant amplitudes for the metastable states are plotted as the others are all zero.}
\end{figure}

\begin{figure}[h!]
\scalebox{0.45}{\includegraphics{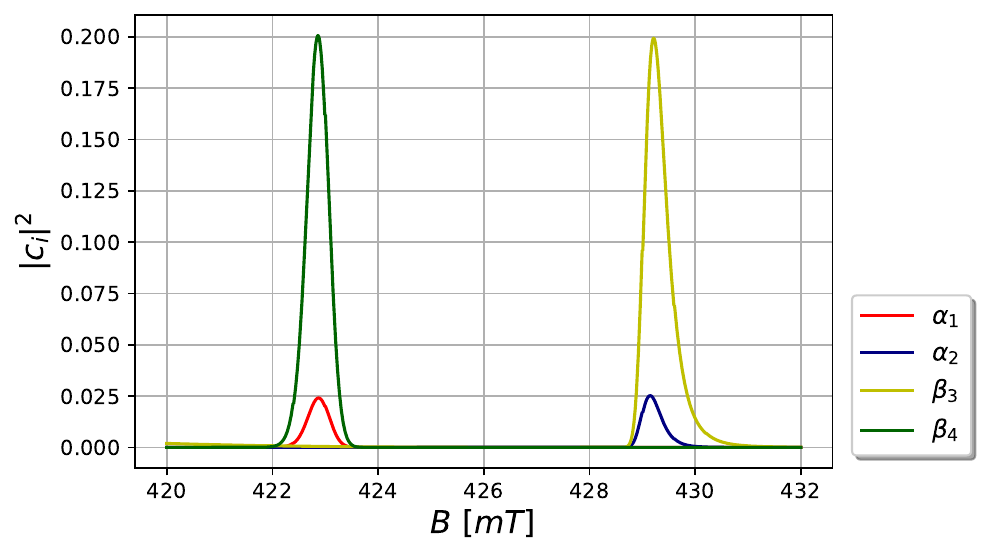}}
\caption{\label{SF2} The probabilities are given as a function of the magnetic field, where its larger values requires a different electromagnetic wave. The frequency used was $f_2=11.94059$~GHz for the mode $\text{TM}_{0,6,0}$ with a field strength of $\varepsilon_0=0.9\cdot10^3\,\frac{\text{V}}{\text{m}}$. Finally, the magnitude of the static electric field was raised to $\varepsilon_x=2.6\cdot10^3\,\frac{\text{V}}{\text{m}}$. Also here, only the four relevant amplitudes are illustrated.}
\end{figure}

To understand the system dynamics it is necessary to solve the Schrödinger equation. The only time-dependent part of all potentials entering the Hamiltonian is the electromagnetic wave of the $\text{TM}_{0,n,0}$ mode. As an analytical solution for the Schrödinger equation is not obtainable, one makes use of the time-dependent perturbation theory~\cite{QM2}. By means of the interaction picture it is possible to write the state $\ket{\psi}$ as a linear combination of the eigenstates of the unperturbed system
\begin{equation}
\ket{\psi(t)}=e^{-\frac{iH_0t}{\hbar}}\ket{\psi(t=t_0)}=\sum_n\underbrace{\braket{n|U(t)|\psi(t=t_0)}}_{=c_n(t)}\ket{n},
\end{equation}
with $U(t)=e^{-\frac{iH_0t}{\hbar}}$ being the time evolution operator in the interaction picture~\cite{QM,QM2}. Correspondingly, it is possible with this state to transform the Schrödinger equation into a set of coupled differential equations for the amplitudes $c_n(t)$
\begin{equation}
i\hbar \dot{c}_k(t)=\sum_n e^{-\frac{i\Delta E t}{\hbar}}c_n(t)\braket{k|V(t)|n},
\end{equation}
with $\Delta E = E_n-E_k$.
In the next steps a short introduction into the time unperturbed system as well as into the perturbating potentials is given. As mentioned above the spin filter contains a static magnetic field pointing in beam direction $\Vec{B}=B_0\hat{e}_z$, whose interaction with the hydrogen atom is described by Eq.~\eqref{BR-Ham}. Its solution for the three sets of $2S_{\nicefrac{1}{2}}$, $2P_{\nicefrac{1}{2}}$ and $2P_{\nicefrac{3}{2}}$ defines the set of unperturbed eigenstates. The eigenenergies depending on the magnetic field amplitude $B_0$ together with their fine structure~\cite{QM2} and Lamb-shift~\cite{Lamb_shift} corrections are visualized in Fig.~\ref{BR}.
\FloatBarrier
To justify the negligence of other states a rough estimated of the energy differences between neighboring states is made. As the interaction is at $B_0\approx0.5$~T, the estimation for the energy difference is done at this point
\begin{eqnarray}
    &&\Delta E_1=E_{\beta_{4}\left(2S_{\nicefrac{1}{2}}\right)}-E_{\alpha_{1}\left(1S_{\nicefrac{1}{2}}\right)}\approx10.2\,\text{eV}\nonumber\\
    &&\Delta E_2=E_{\beta_{4}\left(3S_{\nicefrac{1}{2}}\right)}-E_{g_{1}\left(2P_{\nicefrac{3}{2}}\right)}\approx1.89\,\text{eV}.
\end{eqnarray}
The groundstate, which is of great importance because of its large occupation number, has an energy gap of about $10$ eV to the nearest metastable state, and even the set with $n=3$ is far away with a gap of about $2$ eV. 
Subsequently, the entire Hamiltonian of the system needs to be split as follows
\begin{eqnarray}\label{total_H}
H_{total}&&=\underbrace{H_{\nicefrac{FS}{Lamb}}+H_{BR}}_{=H_0}\nonumber\\&&+\underbrace{V_{Life}+V_{Stark}+V_{RF_{electric}}(t)+V_{RF_{magnetic}}(t)}_{=V(t)}.    
\end{eqnarray}
Starting with the term describing the lifetime by a damping factor, the potential takes the following form
\begin{equation}\label{Life}
    V_{Life}=-i\frac{\hbar}{2}\left(\gamma_1\delta_{l,0}+\gamma_2\delta_{l,1}\right)\delta_{J,J^{\prime}}\delta_{F,F^{\prime}}\delta_{m_F,m_{F^{\prime}}}\mathds{1},
\end{equation}
where $\gamma_i=\frac{1}{\tau_i}$ and $\tau_i$ symbolizes the lifetime of the single states~\cite{QM,QM2}. As each state of the same set has the same lifetime, this reduces the problem to three independent lifetimes. It is also important to note that the lifetime of the $2S$ states is much larger than for the $2P$ states, i.e $\tau_{2S_{\nicefrac{1}{2}}}\gg \tau_{2P_{\nicefrac{1}{2}}}\approx\tau_{2P_{\nicefrac{3}{2}}}$.
\newline
The contribution of the potential describing the interaction of the static electric field is included as a first-order dipole Stark interaction
\begin{equation}\label{Stark_eq}
    V_{Stark}=e\vec{\varepsilon}\cdot\vec{r},
\end{equation}
where $\vec{r}$ is the position operator for the electron and $\vec{\varepsilon}$ the electric field~\cite{Stark}. Before having a closer look at the electromagnetic wave the evolution of the single lifetimes including $H_0$, $V_{Life}$ and $V_{Stark}$ for $\vec{\varepsilon}=\varepsilon_0\hat{e}_x$ is given in Fig.~\ref{Lifetimeplot}.
\FloatBarrier
Its solution is obtained by solving the eigenproblem for those potentials with the necessary parameter given in table~\ref{par_H}. The real part of the eigenvalue corresponds to the eigenenergy, whereas the imaginary part is equal to $-\frac{\Gamma\hbar}{2}=-\frac{\hbar}{2T}$, with $T$ being the new lifetime of the state. Here it should be noted that the lifetime of individual metastable states can be drastically reduced due to the increasing transition probabilities by decreasing energy gaps. Their decay runs over the $2P$ sets, so that the occupation numbers of these short-living $2P$ states rise sharply for a brief moment before decaying into the groundstate. Taking advantage of this moment, it is then possible to populate with the electromagnetic wave the other metastable states that are not strongly affected by the changes in lifetimes.
\newline
The interaction of the electromagnetic wave is described for the electric part similar to the static case by a Stark dipole transition~\eqref{Stark_eq}. The same holds true for the magnetic part described by the interaction given in Eq.~\eqref{BR-Ham} without the hyperfine splitting term. Therefore, only the fields of the $\text{TM}_{0,n,0}$ mode are given here by~\cite{RF}
\begin{eqnarray}\label{RF_equation}
    &&\vec{\varepsilon}=Re\left[\varepsilon_0J_0\left(\frac{\omega_{0,n,0}\rho}{c}\right)e^{i\omega_{0,n,0}t}\right]\nonumber\\&&\vec{B}=Im\left[-\frac{\varepsilon_0}{c}J_1\left(\frac{\omega_{0,n,0}\rho}{c}\right)e^{i\omega_{0,n,0}t}\right],
\end{eqnarray}
with
\begin{equation}
    \omega_{0,n,0}=2\pi f_{0,n,0}=\frac{x_{0,n}c}{R}.
\end{equation}
$f_{0,n,0}$ represents the resonance frequency and $x_{0,n}$ the n-th root of the Bessel function $J_0(x)$.
Therefore, all interactions are given and the system of coupled differential equations can be solved for a pure state. In case of a non-pure state, especially in the case of an unpolarized initial beam condition, the formalism of the density operator $\rho$~\cite{QM} needs to be taken into account. Its time evolution in terms of dissipation is described by the Lindbladian equation~\cite{Lindblad} 
\begin{equation}\label{Lind}
\begin{split}
    \dot{\rho}(t)&=-\frac{i}{\hbar}\left[H(t),\rho(t)\right]\\&+\sum_{i=1}^{16}\gamma_i\left(\sigma_i\rho(t)\sigma^{\dagger}_i-\frac{1}{2}\left\{\sigma^{\dagger}_i\sigma_i,\rho(t)\right\}\right).\\
    \end{split}
\end{equation}
Here the damping factors $\gamma_i$ represent the inverse of the lifetimes introduced in Eq.~\eqref{Life} while the sum runs over all sixteen states. The ladder operators are then defined as 
\begin{equation}
    \sigma_i=\ket{g}\bra{i}\quad\text{and}\quad\sigma^{\dagger}_i=\ket{i}\bra{g},
\end{equation}
where $g$ represents an artificial hydrogen ground state. The Hamiltonian used in the commutator modified slightly from the one given in Eq.~\eqref{total_H} by removing the damping potential $V_{Life}$, as this process is fulfilled by the additional terms behind the commutator in Eq.~\eqref{Lind}. Finally, the last expression symbolizes the anti-commutator $\left\{A,B\right\}=AB+BA$. Only the hyperfine constant has been self-calculated theoretically to $A_{2P_{\nicefrac{3}{2}}}\approx11.84$~MHz. More details are given in the appendix~\ref{A}. The results for a beam passing in the center at $r=0$~m at a velocity of $v=3.095\cdot10^{5}\,\frac{\text{m}}{\text{s}}$ are shown in Fig.~\ref{SF1} and Fig.~\ref{SF2}.
\begin{table}
\caption{\label{par_H} The table gives all necessary values for the parameters to reproduce the simulations given in Fig.~\ref{SF1} and Fig.~\ref{SF2}.}
\begin{ruledtabular}
\begin{tabular}{cccccc}
&$g_j$&$g_I$&$A$~$\left[\text{MHz}\right]$&$\tau$&$\Delta E$~$\left[\text{MHz}\right]$\\
\hline
$2S_{\nicefrac{1}{2}}$&$2.002$~\cite{g_s_neu}&$5.586$~\cite{g_p_neu}&$177.57$~\cite{2SA}&$\frac{1}{8.23}$~s~\cite{Lifetime2S}&\\
$2P_{\nicefrac{1}{2}}$&$0.666$~\footnotemark[2]&$5.586$~\cite{g_p_neu}&$59.22$~\cite{2PA}&$1.6$~ns~\cite{Lifetime2P}&$1057.84$~\cite{Lamb}\footnotemark[1]\\
$2P_{\nicefrac{3}{2}}$&$1.334$~\footnotemark[2]&$5.586$~\cite{g_p_neu}&$11.84$~\cite{2PA}&$1.6$~ns~\cite{Lifetime2P}&$9912.2$~\cite{FS}\footnotemark[3]\\
\end{tabular}
\end{ruledtabular}
\footnotetext[1]{Corresponds to the Lamb-shift.}
\footnotetext[2]{See the formula for the Landé factor in the appendix.}
\footnotetext[3]{Corresponds to the fine splitting.}

\end{table}
\FloatBarrier
\FloatBarrier
The results show that depending on the setting of the magnetic field amplitude mainly one of the four metastable $2S_{\nicefrac{1}{2}}$ states survives the time of flight through the device. To reduce the occupation number of the undesired states as background the static electric field can still be applied along a short distance at the end of the cavity. This then leads to quick decays into the groundstate caused by the much shorter lifetime of the background state.

\subsection{Hydrogen isotopes}
\begin{figure}[h!]
\scalebox{0.45}{\includegraphics{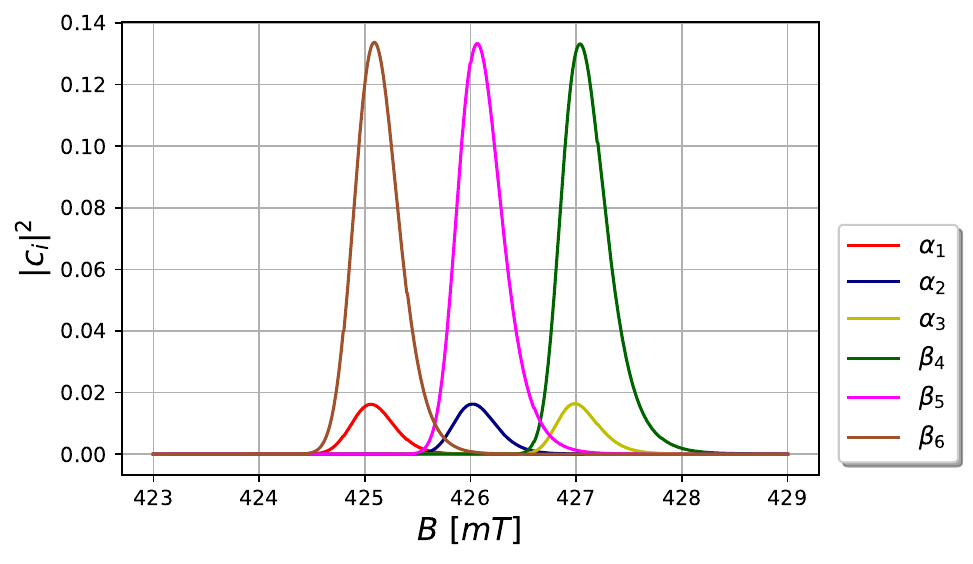}}
\caption{\label{Bild} The absolute value squared of the single amplitudes of the six metastable deuterium states are given as function of the magnetic field. The frequency used was $f_2=11.94059$~GHz for the mode $\text{TM}_{0,6,0}$ with an amplitude of $\varepsilon_0=0.9\cdot10^3\,\frac{\text{V}}{\text{m}}$. In addition, the magnitude of the static electric field is $\varepsilon_x=2.6\cdot10^3\,\frac{\text{V}}{\text{m}}$.}
\end{figure}

So far the focus was onto the hydrogen atom. But also its isotopes deuterium and tritium profit from this device. While tritium has a nuclear spin of $I=\nicefrac{1}{2}$, same as for hydrogen, deuterium has a nuclear spin of $I=1$. Therefore, tritium has the same quantum numbers as hydrogen and only the hyperfine constants differ slightly. This results in a similar behavior for tritium then for hydrogen.
\newline
Deuterium on the other hand has instead of four six metastable states which need to be resolved separately. The necessary solutions for the Breit-Rabi eigenproblem represented in Eq.~\ref{BR-Ham} is given in the appendix~\ref{Deut-BR}. In addition, the dependencies for each state on the different quantum numbers for a vanishing external magnetic field are given in table~\ref{tab_States_D}. The three $\alpha$ states are then separated in the low magnetic field region, whereas the $\beta$ states can be isolated at magnetic fields around $B\approx428$~mT. This is visualized in the following simulation~\ref{Bild}.
\FloatBarrier
The necessary constants which differ from the one of hydrogen are given in table~\ref{par_D}. The frequencies given in the references~\cite{A_S_D,Liste_const}, used for the hyperfine constants, represent the energy differences caused by the hyperfine splitting. From there on one must derive the corresponding value for the hyperfine constant via the Breit-Rabi formulas given in the appendix~\ref{Deut-BR} which results in the prefactor of $\frac{2}{3}$ instead of $1$ in case of hydrogen. The hyperfine constant for the $2P_{\nicefrac{3}{2}}$ set changes slightly for different states due to several different effects. As the results in this paper are obtained for large fields at which the hyperfine splitting is less relevant one simplifies this problem by using the one value for the hyperfine constant given above and especially neglect off-diagonal elements.   
\begin{table}
\caption{\label{tab_States_D} The table defines the Breit-Rabi states at $n=2$ for deuterium in the case of an vanishing external magnetic field $B_0=0$.}
\begin{ruledtabular}
\begin{tabular}{ccccc}
 &$F$&$m_F$&$J$
 &$L$\\
\hline
$\alpha_1$& $\nicefrac{3}{2}$ & $\nicefrac{3}{2}$ & $\nicefrac{1}{2}$ & $0$\\
$\alpha_2$& $\nicefrac{3}{2}$ & $\nicefrac{1}{2}$ & $\nicefrac{1}{2}$ & $0$\\
$\alpha_3$& $\nicefrac{3}{2}$ & $-\nicefrac{1}{2}$ & $\nicefrac{1}{2}$ & $0$\\
$\beta_4$& $\nicefrac{3}{2}$ & $-\nicefrac{3}{2}$ & $\nicefrac{1}{2}$ & $0$\\
$\beta_5$& $\nicefrac{1}{2}$ & $-\nicefrac{1}{2}$ & $\nicefrac{1}{2}$ & $0$\\
$\beta_6$& $\nicefrac{1}{2}$ & $\nicefrac{1}{2}$ & $\nicefrac{1}{2}$ & $0$\\
$e_1$& $\nicefrac{3}{2}$ & $\nicefrac{3}{2}$ & $\nicefrac{1}{2}$ & $1$\\
$e_2$& $\nicefrac{3}{2}$ & $\nicefrac{1}{2}$ & $\nicefrac{1}{2}$ & $1$\\
$e_3$& $\nicefrac{3}{2}$ & $-\nicefrac{1}{2}$ & $\nicefrac{1}{2}$ & $1$\\
$f_4$& $\nicefrac{3}{2}$ & $-\nicefrac{3}{2}$ & $\nicefrac{1}{2}$ & $1$\\
$f_5$& $\nicefrac{1}{2}$ & $-\nicefrac{1}{2}$ & $\nicefrac{1}{2}$ & $1$\\
$f_6$& $\nicefrac{1}{2}$ & $\nicefrac{1}{2}$ & $\nicefrac{1}{2}$ & $1$\\
$g_1$& $\nicefrac{5}{2}$ & $\nicefrac{5}{2}$ & $\nicefrac{3}{2}$ & $1$\\
$g_2$& $\nicefrac{5}{2}$ & $\nicefrac{3}{2}$ & $\nicefrac{3}{2}$ & $1$\\
$g_3$& $\nicefrac{5}{2}$ & $\nicefrac{1}{2}$ & $\nicefrac{3}{2}$ & $1$\\
$g_4$& $\nicefrac{5}{2}$ & $-\nicefrac{1}{2}$ & $\nicefrac{3}{2}$ & $1$\\
$g_5$& $\nicefrac{5}{2}$ & $-\nicefrac{3}{2}$ & $\nicefrac{3}{2}$ & $1$\\
$g_6$& $\nicefrac{5}{2}$ & $-\nicefrac{5}{2}$ & $\nicefrac{3}{2}$ & $1$\\
$h_7$& $\nicefrac{3}{2}$ & $\nicefrac{3}{2}$ & $\nicefrac{3}{2}$ & $1$\\
$h_8$& $\nicefrac{3}{2}$ & $\nicefrac{1}{2}$ & $\nicefrac{3}{2}$ & $1$\\
$h_9$& $\nicefrac{3}{2}$ & $-\nicefrac{1}{2}$ & $\nicefrac{3}{2}$ & $1$\\
$h_{10}$& $\nicefrac{3}{2}$ & $-\nicefrac{3}{2}$ & $\nicefrac{3}{2}$ & $1$\\
$k_{11}$& $\nicefrac{1}{2}$ & $\nicefrac{1}{2}$ & $\nicefrac{3}{2}$ & $1$\\
$k_{12}$& $\nicefrac{1}{2}$ & $-\nicefrac{1}{2}$ & $\nicefrac{3}{2}$ & $1$\\
\end{tabular}
\end{ruledtabular}
\end{table}
\begin{table}
\caption{\label{par_D} In addition to the unchanged parameters of $\tau$ and $g_j$, which can be found in table~\ref{par_H}, this table gives all necessary values for the parameters to reproduce the simulation given in Fig.~\ref{Bild}.}
\begin{ruledtabular}
\begin{tabular}{cccc}
&$g_I$&$A$~$\left[\text{MHz}\right]$&$\Delta E$~$\left[\text{MHz}\right]$\\
\hline
$2S_{\nicefrac{1}{2}}$&$0.857438$~\cite{g_p_d}&$\frac{2}{3}\cdot40.924454$~\cite{A_S_D}\footnotemark[4]&\\
$2P_{\nicefrac{1}{2}}$&$0.857438$~\cite{g_p_d}&$\frac{2}{3}\cdot13.633390$~\cite{Liste_const}\footnotemark[4]&$1058.49$~\cite{Lamb_D}\footnotemark[1]\\
$2P_{\nicefrac{3}{2}}$&$0.857438$~\cite{g_p_d}&$1.817671$~\cite{Liste_const}\footnotemark[4]&$9912.59$~\cite{FS_D}\footnotemark[3]\\
\end{tabular}
\end{ruledtabular}
\footnotetext[1]{Corresponds to the Lamb-shift.}
\footnotetext[2]{See the formula for the Landé factor in the appendix.}
\footnotetext[3]{Corresponds to the fine splitting.}
\footnotetext[4]{Deviations from the values given in the sources are explained above in the text}
\end{table}
\section{Experimental details}
To exploit the predictions of the theory described before the two main parts of the device, the magnetic field configuration as well as the cavity need to be discussed.

\subsection{Magnetic field configuration}
The purpose of the magnetic field configuration is to achieve a homogeneous constant magnetic field in beam direction over the distance of the length of the cavity. 
For the realization of a magnetic field in longitudinal direction of about $426$~mT, but also $57$~mT, several options have been considered. The most straightforward method would be to use a solenoid with a length of about $500$~mm and an inner diameter determined by the outer diameter of the cavity of about $150$~mm. The obvious advantage would be that the field can easily be adjusted to these required values, but an estimate shows that the electric power requirement will be on the order of $10$~kW, which would require the use of a dedicated cooling system. We discarded this option because of this prospect. On the other hand a system of Halbach magnet rings would be able to generate a static field of $400$~mT, but the switching to the lower field value would require a complicated mechanical setup, which would rotate the Halbach rings against each other. 
Therefore, this option was discarded as well. At present we favor the realization by means of an array of superconducting solenoids, for which we have a first design. This design features a layer of larger superconducting solenoids, which serves to reduce the stray magnetic field, but avoids zero crossings of the longitudinal field.
\newline
Up to now the longitudinal magnetic field $B_z$ and its conditions around the cavity have been discussed, but let us draw the attention towards the radial magnetic field component $B_{\rho}$ as well as the conditions outside of the cavity. For cylindrical magnetic field shapes the relation between the longitudinal field and the radial component can be derived directly from Maxwell's equation $\vec{\nabla}\cdot \vec{B}=0$ to
\begin{equation}
    B_{\rho}=-\frac{\rho}{2}\frac{\partial B_z}{\partial z}.
\end{equation}
Consequently, the radial component vanishes in the part around the cavity which is intended. Nevertheless, the gradients, necessary to achieve the homogeneous magnetic field in the center around the cavity, can produce electric fields in the rest frame of those atoms positioned off-axis, which then quenches them to the groundstate. Therefore, for special applications with a large count rate one would need to have a long holding field and by this a slow decrease over length to reduce the electric field seen in the rest frame by the atoms. 

\subsection{Cavity}
\begin{figure}[h!]
\scalebox{0.45}{\includegraphics{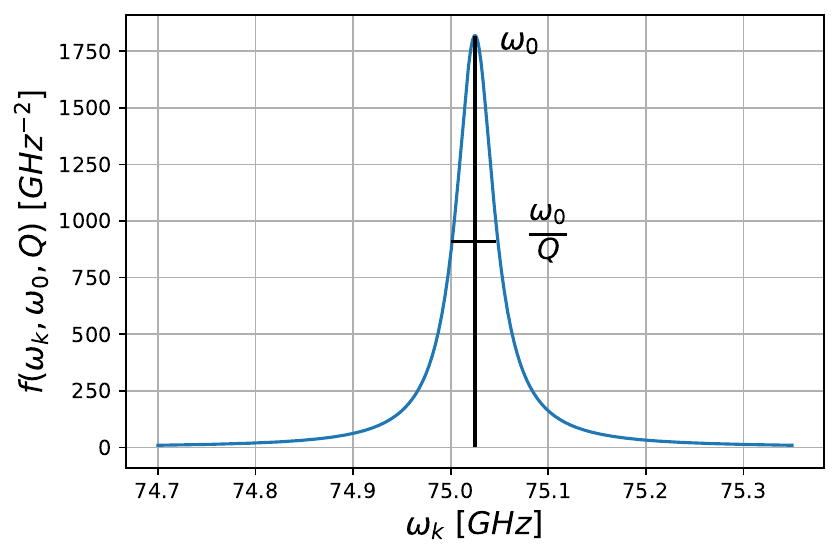}}
\caption{\label{Lorentz}The Lorentz curve is given for a resonance frequency of $f_0\approx 11.94059$~GHz and a quality factor of $Q=1600$.}
\end{figure}

\begin{figure}[h!]
\scalebox{0.45}{\includegraphics{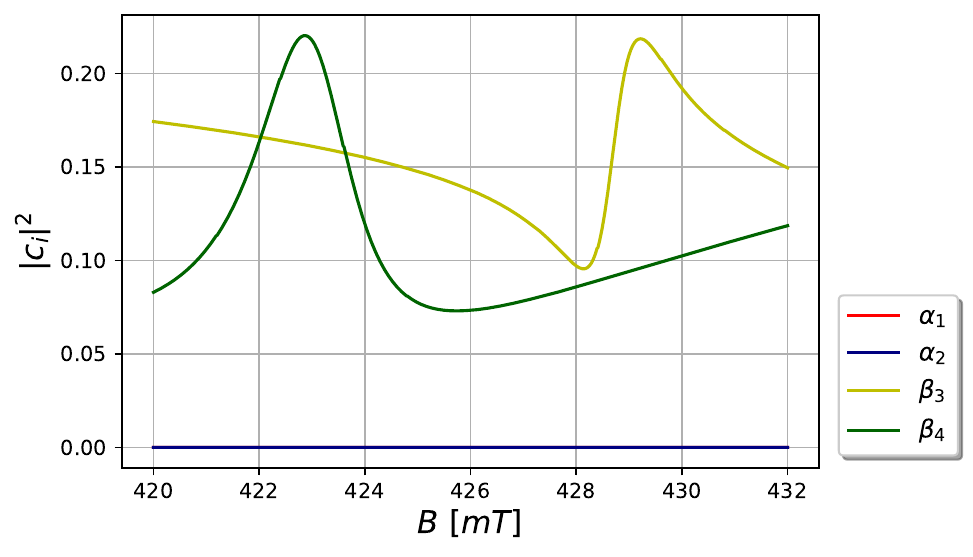}}
\caption{\label{Quality_Q1600}Simulation of the probabilities to find the single $n=2$ states for a cavity given with a quality factor of $Q=1600$.}
\end{figure}

\begin{figure}[h!]
\scalebox{0.45}{\includegraphics{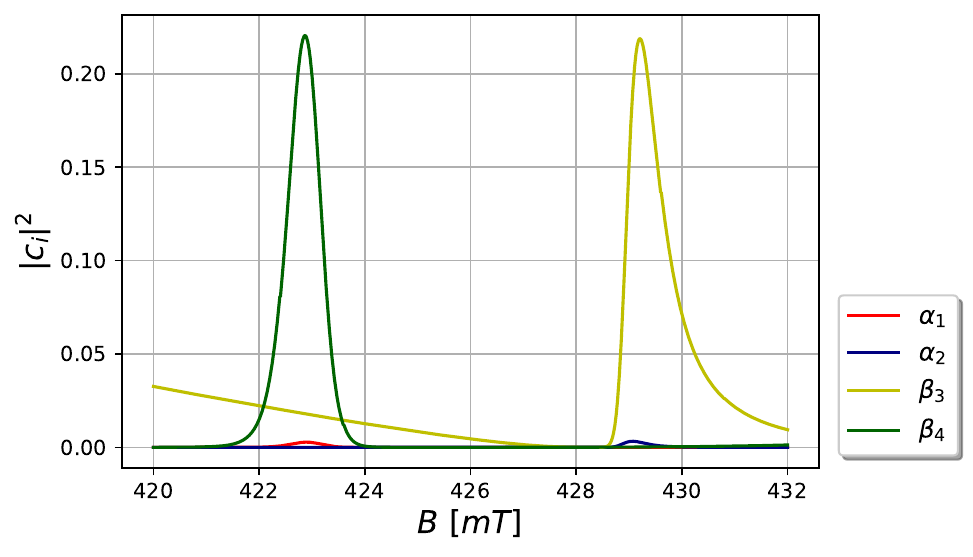}}
\caption{\label{Quality_Q16000}Simulation of the probabilities to find the single $n=2$ states for a cavity given with a quality factor of $Q=16000$.}
\end{figure}

\begin{figure}[h!]
\scalebox{0.45}{\includegraphics{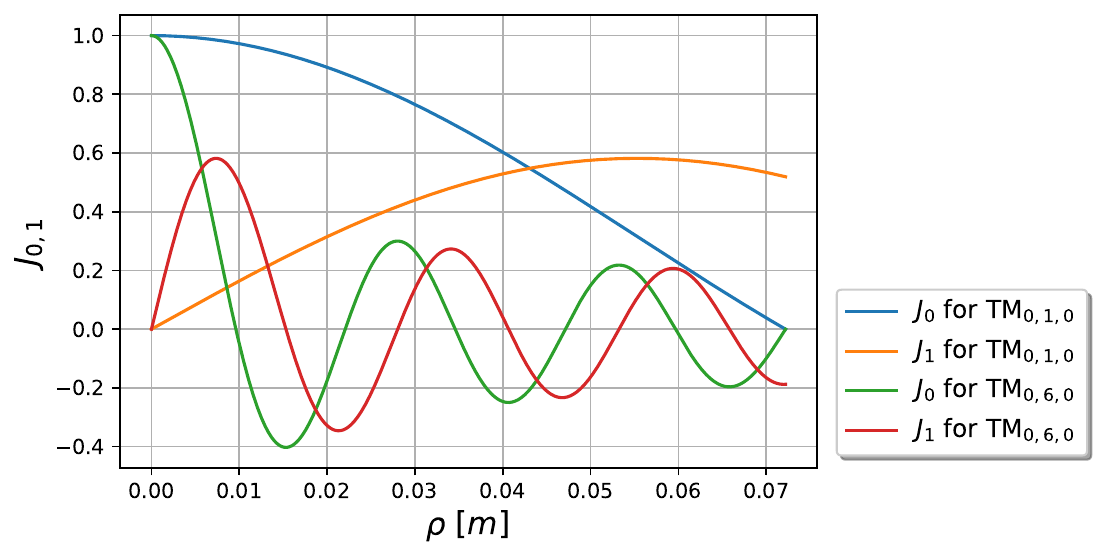}}
\caption{\label{Bessel}The Bessel functions are illustrated for the two transversal magnetic modes $\text{TM}_{0,1/6,0}$ with respect to the running parameter $\rho$ of the radius $R$ for the cavity.}
\end{figure}

As described in reference~\cite{Ralf}, the cavity consists of four quadrants forming a cylinder with the height $h$ and the radius $R$. On two parallel ones the static electric field is applied while on the others the electromagnetic waves are induced and the power output is measured. In addition, the cavity should be able to create both resonance frequencies and has a sufficiently high quality factor $Q$ to resolve the individual peaks. Moreover, it needs to provide the coupled electromagnetic waves with enough power to make transitions possible. As the quadrants are isolated, meaning not connected to each other, the quality factor is much worse than theoretically expected. For the already existing device a quality factor between $1000$ and $3000$ is sufficient to fulfil the conditions permitting the separation of the two $\alpha$ peaks from each other~\cite{Ralf}. The quality factor $Q$ is defined by
\begin{equation}
    Q=\frac{f_0}{\Delta f},
\end{equation}
with $f_0$ being the resonance frequency and $\Delta f$ the frequency half width. The resonance frequency of the $\text{TM}_{0,6,0}$ mode is roughly ten times larger compared to the $\text{TM}_{0,1,0}$. If one wants to run the new resonance frequency with the same cavity one would need to realize a quality factor of $Q\approx 16000$.
To prove this, one needs to simulate not only one possible frequency, which is perfectly coupled into the system but a bunch of modes with decreasing power for each mode further away from the resonance. A typical frequency spectrum inside a cavity~\cite{RF} is represented by a Lorentz distribution
\begin{equation}
    f(\omega,\omega_0,Q)=\frac{1}{\left(\omega-\omega_0\right)^2+\left(\frac{\omega_0}{2Q}\right)^2},
\end{equation}
and illustrated in Fig.~\ref{Lorentz}. The maximum of the Lorentz curve is given at the resonance frequency, whereas the half width depends on the quality factor $Q$. In the limit $Q\to \infty$ the Lorentz curve turns into a delta distribution at the point of the resonance frequency. This represents the case of an ideal cavity, which has been discussed in the previous sections.
\FloatBarrier
Moreover, the entire electric field created inside the cavity by the electromagnetic waves can be modeled as a sum over all modes
\begin{equation}
        \vec{\epsilon}_{RF}(t)=\sum_k \vec{\epsilon}_{{RF}_k}(t)\approx \sum_k \epsilon_{0,k} J_0\left(\frac{x_{0,n}\rho}{R}\right) \cos\left(\omega_k t\right)\hat{e}_z. 
\end{equation}
The decrease in power for each mode of resonance is embedded in the amplitudes $\epsilon_{0,k}$ 
\begin{equation}
        \epsilon_{0,k}=\epsilon_0 \frac{f\left(\omega_k,\omega_0,Q\right)}{\sum_k f\left(\omega_k,\omega_0,Q\right)},
\end{equation}
where the denominator is given to normalize the sum of all modes to a given entire electric field amplitude $\epsilon_0$ corresponding to the applied power. Modified simulations show different outcomes depending on the quality factor $Q$ in Fig.~\ref{Quality_Q1600} and Fig.~\ref{Quality_Q16000}. In Fig.~\ref{Quality_Q1600} a rather small quality factor $Q=1600$ is used which allows too many modes to couple into the cavity such that the peaks cannot be properly separated from each other. In the case of a larger quality factor $Q=16000$, in Fig.~\ref{Quality_Q16000}, the background is reduced and the peaks take similar shapes as in the case of an ideal cavity.
\FloatBarrier
\FloatBarrier
Let us have a closer look on the electric fields from the electromagnetic waves produced in the cavity. As they are directly proportional to the Bessel functions $J_0$ represented in Fig.~\ref{Bessel} it is enough to focus on them. 
\FloatBarrier
A large difference between the two modes is visible in the Bessel functions in Fig.~\ref{Bessel}. The beam enters the cavity with a given beam profile and, more importantly, a non-negligible beam profile in the diameter. For the $\text{TM}_{0,1,0}$ mode this is not very important as the Bessel function $J_{0\left(\text{TM}_{0,1,0}\right)}$ decreases slowly. In contrast the Bessel function of mode $\text{TM}_{0,6,0}$ consist of six roots meaning that $J_{0\left(\text{TM}_{0,6,0}\right)}$ varies much faster. If the beam diameter exceeds $0.5$~cm parts of the $\beta_{3,4}$ states for the first respectively second peak would survive and produce an additional background. Therefore, the separation of both peaks would get worse. This problem can be avoided by an aperture in front of the device to reduce the beam size. As discussed previously this is an idea to realize both spin filter types in one single device. Depending on the experiment a different mode or cavity type could be employed to avoid the smaller count rate caused by using an aperture.\newline
As mentioned above the radius $R$ of the cavity needs to be changed for operating the two different modes. Given that $R_2>R_1$ the mode $\text{TM}_{0,1,0}$ needs to be tuned by $\Delta R$. This can be realized by introducing small bars equally distributed from the surface inside the cavity controlled by a motor.

\newpage
\section{Conclusion and Outlook}
This new device combines the previous known method to separate single $\alpha$ states from a metastable hydrogen beam with the ability to filter also individual $\beta$ states. Not only works the device for hydrogen but also for its isotopes, tritium and deuterium. As the analysis of the Schrödinger equation is very accurate in describing the kinematics of a system with the new spin filter, more states can be verified experimentally. One of the experiments that benefits from this is the spectroscopy measurements involving a special configuration of a Sona transition unit~\cite{engels2021}. As in this experiment quantum oscillations between the four metastable hydrogen $2S_{\nicefrac{1}{2}}$ states are visible but currently the occupation numbers of the two $\alpha$ states individually are measurable. When building such a device, care must be taken to ensure that the magnetic field is very homogeneous at the point where the cavity is installed, as any kind of disturbance would minimize the intensity. Furthermore, the design of the cavity as the central component can be challenging if one wants to combine both types of spin filters in one device.
\newline
To conclude, also the helium ion and its prominent isotope $^3\text{He}^+$ are interesting candidates to profit from this development as their metastable states have reasonable long lifetimes~\cite{Lifetime_He} and their ionic structures are hydrogenlike. Therefore, it should be possible to extend the spin filter concept for their purpose.

\begin{acknowledgments}
We wish to thank Prof.~J.~Pretz and Prof.~C.~Hanhart for their supervision and help to realize this paper. In addition, C. Kannis acknowledges funding from the Deutsche Forschungsgemeinschaft (DFG, German Research Foundation) – 533904660.
\end{acknowledgments}

\appendix

\section{Breit-Rabi solutions for hydrogen}\label{BR-States}
For this paper a lot of states and their eigenenergies have been introduced. In this section these Breit-Rabi states and their energies fulfilling the eigenproblem of Eq.~\eqref{BR-Ham} in addition to their relative energy corrections due to fine splitting $\left(FS\right)$ and Lamb shift $\left(\Delta E_{Lamb}\right)$ are defined. First for the $2S_{\nicefrac{1}{2}}$ set in the $\Ket{F,m_F,J,L}$ basis
\begin{eqnarray}
    &&\ket{\alpha_1}=\Ket{1,1,\nicefrac{1}{2},0}\nonumber\\
    &&E_{\alpha_1}=\frac{A_{2S_{\nicefrac{1}{2}}}}{4}+\frac{1}{2}\left(g_S\mu_B-g_I\mu_k\right)B_0\nonumber\\
    &&\ket{\alpha_2}=\frac{1}{\sqrt{1+x^2(B)}}\left(x(B)\ket{0,0,\nicefrac{1}{2},0}+\ket{1,0,\nicefrac{1}{2},0}\right)\nonumber\\
    &&\text{with}\quad x(B)=\frac{\left(g_S\mu_B+g_I\mu_k\right)B_0}{A_{2S_{\nicefrac{1}{2}}}+\sqrt{A^2_{2S_{\nicefrac{1}{2}}}+\left(g_S\mu_B+g_I\mu_k\right)^2B_0^2}}\nonumber\\
    &&E_{\alpha_2}=-\frac{A_{2S_{\nicefrac{1}{2}}}}{4}+\frac{1}{2}\sqrt{A^2_{2S_{\nicefrac{1}{2}}}+\left(g_S\mu_B+g_I\mu_k\right)^2B_0^2}\nonumber\\
    &&\ket{\beta_3}=\ket{1,-1,\nicefrac{1}{2},0}\\
    &&E_{\beta_3}=\frac{A_{2S_{\nicefrac{1}{2}}}}{4}-\frac{1}{2}\left(g_S\mu_B-g_I\mu_k\right)B_0\nonumber\\
    &&\ket{\beta_4}=\frac{1}{\sqrt{1+\omega^2(B)}}\left(\ket{0,0,\nicefrac{1}{2},0}+\omega(B)\ket{1,0,\nicefrac{1}{2},0}\right)\nonumber\\
    &&\text{with}\quad \omega(B)=\frac{A_{2S_{\nicefrac{1}{2}}}-\sqrt{A^2_{2S_{\nicefrac{1}{2}}}+\left(g_S\mu_B+g_I\mu_k\right)^2B_0^2}}{\left(g_S\mu_B+g_I\mu_k\right)B_0}\nonumber\\
    &&E_{\beta_4}=-\frac{A_{2S_{\nicefrac{1}{2}}}}{4}-\frac{1}{2}\sqrt{A^2_{2S_{\nicefrac{1}{2}}}+\left(g_S\mu_B+g_I\mu_k\right)^2B_0^2}\nonumber.
\end{eqnarray}
Analogue for $2P_{\nicefrac{1}{2}}$
\begin{eqnarray}
    &&\ket{e_1}=\ket{1,1,\nicefrac{1}{2},1}\nonumber\\
    &&E_{e_1}=-\Delta E_{Lamb}+\frac{A_{2P_{\nicefrac{1}{2}}}}{4}+\frac{1}{2}\left(g_{J_{2P_{\nicefrac{1}{2}}}}\mu_B-g_I\mu_k\right)B_0\nonumber\\
    &&\ket{e_2}=\frac{1}{\sqrt{1+y^2(B)}}\left(y(B)\ket{0,0,\nicefrac{1}{2},1}+\ket{1,0,\nicefrac{1}{2},1}\right)\nonumber\\
    &&\text{with}\quad y(B)=\frac{\left(g_{J_{2P_{\nicefrac{1}{2}}}}\mu_B+g_I\mu_k\right)B_0}{A_{2P_{\nicefrac{1}{2}}}+\sqrt{A^2_{2P_{\nicefrac{1}{2}}}+\left(g_{J_{2P_{\nicefrac{1}{2}}}}\mu_B+g_I\mu_k\right)^2B_0^2}}\nonumber\\
    &&E_{e_2}=-\Delta E_{Lamb}-\frac{A_{2P_{\nicefrac{1}{2}}}}{4}+\frac{1}{2}\sqrt{A^2_{2P_{\nicefrac{1}{2}}}+\left(g_{J_{2P_{\nicefrac{1}{2}}}}\mu_B+g_I\mu_k\right)^2B_0^2}\nonumber\\
    &&\ket{f_3}=\ket{1,-1,\nicefrac{1}{2},1}\\
    &&E_{f_3}=-\Delta E_{Lamb}+\frac{A_{2P_{\nicefrac{1}{2}}}}{4}-\frac{1}{2}\left(g_{J_{2P_{\nicefrac{1}{2}}}}\mu_B-g_I\mu_k\right)B_0\nonumber\\
    &&\ket{f_4}=\frac{1}{\sqrt{1+z^2(B)}}\left(\ket{0,0,\nicefrac{1}{2},1}+z(B)\ket{1,0,\nicefrac{1}{2},1}\right)\nonumber\\
    &&\text{with}\quad z(B)=\frac{A_{2P_{\nicefrac{1}{2}}}-\sqrt{A^2_{2P_{\nicefrac{1}{2}}}+\left(g_{J_{2P_{\nicefrac{1}{2}}}}\mu_B+g_I\mu_k\right)^2B_0^2}}{\left(g_{J_{2P_{\nicefrac{1}{2}}}}\mu_B+g_I\mu_k\right)B_0}\nonumber\\
    &&E_{f_4}=-\Delta E_{Lamb}-\frac{A_{2P_{\nicefrac{1}{2}}}}{4}-\frac{1}{2}\sqrt{A^2_{2P_{\nicefrac{1}{2}}}+\left(g_{J_{2P_{\nicefrac{1}{2}}}}\mu_B+g_I\mu_k\right)^2B_0^2}\nonumber.
\end{eqnarray}
\newpage
Finally, for the $2P_{\nicefrac{3}{2}}$ set
\begin{widetext}
\begin{eqnarray}
    &&\ket{g_1}=\ket{2,2,\nicefrac{3}{2},1}\nonumber\\
    &&E_{g_1}=FS+\frac{3A_{2P_{\nicefrac{3}{2}}}}{4}+\frac{1}{2}\left(3g_{J_{2P_{\nicefrac{3}{2}}}}\mu_B-g_I\mu_k\right)B_0\nonumber\\
    &&\ket{g_2}=\frac{1}{\sqrt{1+\chi_1^2(B)}}\left(\ket{2,1,\nicefrac{3}{2},1}+\chi_1(B)\ket{1,1,\nicefrac{3}{2},1}\right)\nonumber\\
    &&E_{g_2}=FS-\frac{A_{2P_{\nicefrac{3}{2}}}}{4}+g_{J_{2P_{\nicefrac{3}{2}}}}B_0+\sqrt{A^2_{2P_{\nicefrac{3}{2}}}-\frac{A_{2P_{\nicefrac{3}{2}}}}{2}\left(g_{J_{2P_{\nicefrac{3}{2}}}}\mu_B+g_I\mu_k\right)B_0+\frac{1}{4}\left(g_{J_{2P_{\nicefrac{3}{2}}}}\mu_B+g_I\mu_k\right)^2B_0^2}\nonumber\\
    &&\ket{g_3}=\frac{1}{\sqrt{1+\epsilon_1^2(B)}}\left(\ket{2,0,\nicefrac{3}{2},1}+\epsilon_1(B)\ket{1,0,\nicefrac{3}{2},1}\right)\nonumber\\
    &&E_{g_3}=FS-\frac{A_{2P_{\nicefrac{3}{2}}}}{4}+\sqrt{A^2_{2P_{\nicefrac{3}{2}}}+\frac{1}{4}\left(g_{J_{2P_{\nicefrac{3}{2}}}}\mu_B+g_I\mu_k\right)^2B_0^2}\nonumber\\
    &&\ket{g_4}=\frac{1}{\sqrt{1+\kappa_1^2(B)}}\left(\ket{2,-1,\nicefrac{3}{2},1}+\kappa_1(B)\ket{1,-1,\nicefrac{3}{2},1}\right)\nonumber\\
    &&E_{g_4}=FS-\frac{A_{2P_{\nicefrac{3}{2}}}}{4}-g_{J_{2P_{\nicefrac{3}{2}}}}B_0+\sqrt{A^2_{2P_{\nicefrac{3}{2}}}+\frac{A_{2P_{\nicefrac{3}{2}}}}{2}\left(g_{J_{2P_{\nicefrac{3}{2}}}}\mu_B+g_I\mu_k\right)B_0+\frac{1}{4}\left(g_{J_{2P_{\nicefrac{3}{2}}}}\mu_B+g_I\mu_k\right)^2B_0^2}\nonumber\\
    &&\ket{g_5}=\ket{2,-2,\nicefrac{3}{2},1}\nonumber\\
    &&E_{g_5}=FS-\frac{3A_{2P_{\nicefrac{3}{2}}}}{4}-\frac{1}{2}\left(3g_{J_{2P_{\nicefrac{3}{2}}}}\mu_B-g_I\mu_k\right)B_0\\
    &&\ket{h_6}=\frac{1}{\sqrt{1+\chi_2^2(B)}}\left(\chi_2(B)\ket{2,1,\nicefrac{3}{2},1}+\ket{1,1,\nicefrac{3}{2},1}\right)\nonumber\\
    &&E_{h_6}=FS-\frac{A_{2P_{\nicefrac{3}{2}}}}{4}+g_{J_{2P_{\nicefrac{3}{2}}}}B_0-\sqrt{A^2_{2P_{\nicefrac{3}{2}}}-\frac{A_{2P_{\nicefrac{3}{2}}}}{2}\left(g_{J_{2P_{\nicefrac{3}{2}}}}\mu_B+g_I\mu_k\right)B_0+\frac{1}{4}\left(g_{J_{2P_{\nicefrac{3}{2}}}}\mu_B+g_I\mu_k\right)^2B_0^2}\nonumber\\
    &&\ket{h_7}=\frac{1}{\sqrt{1+\epsilon_2^2(B)}}\left(\epsilon_2(B)\ket{2,0,\nicefrac{3}{2},1}+\ket{1,0,\nicefrac{3}{2},1}\right)\nonumber\\
    &&E_{h_7}=FS-\frac{A_{2P_{\nicefrac{3}{2}}}}{4}-\sqrt{A^2_{2P_{\nicefrac{3}{2}}}+\frac{1}{4}\left(g_{J_{2P_{\nicefrac{3}{2}}}}\mu_B+g_I\mu_k\right)^2B_0^2}\nonumber\\
    &&\ket{h_8}=\frac{1}{\sqrt{1+\kappa_2^2(B)}}\left(\kappa_2(B)\ket{2,-1,\nicefrac{3}{2},1}+\ket{1,-1,\nicefrac{3}{2},1}\right)\nonumber\\
    &&E_{h_8}=FS-\frac{A_{2P_{\nicefrac{3}{2}}}}{4}-g_{J_{2P_{\nicefrac{3}{2}}}}B_0-\sqrt{A^2_{2P_{\nicefrac{3}{2}}}+\frac{A_{2P_{\nicefrac{3}{2}}}}{2}\left(g_{J_{2P_{\nicefrac{3}{2}}}}\mu_B+g_I\mu_k\right)B_0+\frac{1}{4}\left(g_{J_{2P_{\nicefrac{3}{2}}}}\mu_B+g_I\mu_k\right)^2B_0^2}\nonumber\\
    &&\text{with}\quad \chi_{\nicefrac{1}{2}}(B)=-\frac{\sqrt{3}}{4}\frac{\left(g_{J_{2P_{\nicefrac{3}{2}}}}\mu_B+g_I\mu_k\right)B_0}{\mp A_{2P_{\nicefrac{3}{2}}}\pm\frac{1}{4}\left(g_{J_{2P_{\nicefrac{3}{2}}}}\mu_B+g_I\mu_k\right)B_0\mp\sqrt{A^2_{2P_{\nicefrac{3}{2}}}-\frac{A_{2P_{\nicefrac{3}{2}}}}{2}\left(g_{J_{2P_{\nicefrac{3}{2}}}}\mu_B+g_I\mu_k\right)B_0+\frac{1}{4}\left(g_{J_{2P_{\nicefrac{3}{2}}}}\mu_B+g_I\mu_k\right)^2B_0^2}}\nonumber\\
    &&\text{with}\quad \epsilon_{\nicefrac{1}{2}}(B)=-\frac{1}{2}\frac{\left(g_{J_{2P_{\nicefrac{3}{2}}}}\mu_B+g_I\mu_k\right)B_0}{\mp A_{2P_{\nicefrac{3}{2}}}\mp\sqrt{A^2_{2P_{\nicefrac{3}{2}}}+\frac{1}{4}\left(g_{J_{2P_{\nicefrac{3}{2}}}}\mu_B+g_I\mu_k\right)^2B_0^2}}\nonumber\\
    &&\text{with}\quad \kappa_{\nicefrac{1}{2}}(B)=-\frac{\sqrt{3}}{4}\frac{\left(g_{J_{2P_{\nicefrac{3}{2}}}}\mu_B+g_I\mu_k\right)B_0}{\mp A_{2P_{\nicefrac{3}{2}}}\mp\frac{1}{4}\left(g_{J_{2P_{\nicefrac{3}{2}}}}\mu_B+g_I\mu_k\right)B_0\mp\sqrt{A^2_{2P_{\nicefrac{3}{2}}}+\frac{A_{2P_{\nicefrac{3}{2}}}}{2}\left(g_{J_{2P_{\nicefrac{3}{2}}}}\mu_B+g_I\mu_k\right)B_0+\frac{1}{4}\left(g_{J_{2P_{\nicefrac{3}{2}}}}\mu_B+g_I\mu_k\right)^2B_0^2}}\nonumber.
\end{eqnarray}
\end{widetext}
\newpage
\newpage
\section{Breit-Rabi solutions for deuterium}\label{Deut-BR}
Analogously to the above defined hydrogen states and their energys the solutions for Eq.~\eqref{BR-Ham} in the case of a deuterium atom are introduced here. First starting with the set $2S_{\nicefrac{1}{2}}$ again in the $\ket{F,m_F,J,L}$ basis
\begin{widetext}
\begin{eqnarray}
    &&\ket{\alpha_1}=\Ket{\nicefrac{3}{2},\nicefrac{3}{2},\nicefrac{1}{2},0}\nonumber\\
    &&E_{\alpha_1}=\frac{A_{2S_{\nicefrac{1}{2}}}}{2}+\left(\frac{g_S\mu_B}{2}-g_I\mu_k\right)B_0\nonumber\\
    &&\ket{\alpha_2}=\frac{1}{\sqrt{1+\gamma_1^2(B)}}\left(\gamma_1(B)\ket{\nicefrac{1}{2},\nicefrac{1}{2},\nicefrac{1}{2},0}+\ket{\nicefrac{3}{2},\nicefrac{1}{2},\nicefrac{1}{2},0}\right)\nonumber\\
    &&E_{\alpha_2}=-\frac{A_{2S_{\nicefrac{1}{2}}}}{4}-\frac{g_I\mu_kB_0}{2}+\sqrt{\frac{A^2_{2S_{\nicefrac{1}{2}}}}{2}+\left(\frac{A_{2S_{\nicefrac{1}{2}}}}{4}+\frac{1}{2}\left(g_S\mu_B+g_I\mu_k\right)B_0\right)^2}\nonumber\\
    &&\ket{\alpha_3}=\frac{1}{\sqrt{1+\Gamma_1^2(B)}}\left(\Gamma_1(B)\ket{\nicefrac{1}{2},-\nicefrac{1}{2},\nicefrac{1}{2},0}+\ket{\nicefrac{3}{2},-\nicefrac{1}{2},\nicefrac{1}{2},0}\right)\nonumber\\
    &&E_{\alpha_3}=-\frac{A_{2S_{\nicefrac{1}{2}}}}{4}+\frac{g_I\mu_kB_0}{2}+\sqrt{\frac{A^2_{2S_{\nicefrac{1}{2}}}}{2}+\left(\frac{A_{2S_{\nicefrac{1}{2}}}}{4}-\frac{1}{2}\left(g_S\mu_B+g_I\mu_k\right)B_0\right)^2}\nonumber\\
    &&\ket{\beta_4}=\Ket{\nicefrac{3}{2},-\nicefrac{3}{2},\nicefrac{1}{2},0}\nonumber\\
    &&E_{\beta_4}=\frac{A_{2S_{\nicefrac{1}{2}}}}{2}-\left(\frac{g_S\mu_B}{2}-g_I\mu_k\right)B_0\nonumber\\
    &&\ket{\beta_5}=\frac{1}{\sqrt{1+\Gamma_2^2(B)}}\left(\ket{\nicefrac{1}{2},-\nicefrac{1}{2},\nicefrac{1}{2},0}+\Gamma_2(B)\ket{\nicefrac{3}{2},-\nicefrac{1}{2},\nicefrac{1}{2},0}\right)\nonumber\\
    &&E_{\beta_5}=-\frac{A_{2S_{\nicefrac{1}{2}}}}{4}+\frac{g_I\mu_kB_0}{2}-\sqrt{\frac{A^2_{2S_{\nicefrac{1}{2}}}}{2}+\left(\frac{A_{2S_{\nicefrac{1}{2}}}}{4}-\frac{1}{2}\left(g_S\mu_B+g_I\mu_k\right)B_0\right)^2}\nonumber\\
    &&\ket{\beta_6}=\frac{1}{\sqrt{1+\gamma_2^2(B)}}\left(\ket{\nicefrac{1}{2},\nicefrac{1}{2},\nicefrac{1}{2},0}+\gamma_2(B)\ket{\nicefrac{3}{2},\nicefrac{1}{2},\nicefrac{1}{2},0}\right)\nonumber\\
    &&E_{\beta_6}=-\frac{A_{2S_{\nicefrac{1}{2}}}}{4}-\frac{g_I\mu_kB_0}{2}-\sqrt{\frac{A^2_{2S_{\nicefrac{1}{2}}}}{2}+\left(\frac{A_{2S_{\nicefrac{1}{2}}}}{4}+\frac{1}{2}\left(g_S\mu_B+g_I\mu_k\right)B_0\right)^2}\nonumber\\
    &&\text{with}\quad \gamma_{\nicefrac{1}{2}}(B)=\mp\frac{\sqrt{2}}{3}\frac{\left(g_S\mu_B+g_I\mu_k\right)B_0}{\frac{3A_{2S_{\nicefrac{1}{2}}}}{4}+\frac{1}{6}\left(g_S\mu_B+g_I\mu_k\right)B_0+\sqrt{\frac{A^2_{2S_{\nicefrac{1}{2}}}}{2}+\left(\frac{A_{2S_{\nicefrac{1}{2}}}}{4}+\frac{1}{2}\left(g_S\mu_B+g_I\mu_k\right)B_0\right)^2}}\nonumber\\
    &&\text{with}\quad \Gamma_{\nicefrac{1}{2}}(B)=\mp\frac{\sqrt{2}}{3}\frac{\left(g_S\mu_B+g_I\mu_k\right)B_0}{\frac{3A_{2S_{\nicefrac{1}{2}}}}{4}-\frac{1}{6}\left(g_S\mu_B+g_I\mu_k\right)B_0+\sqrt{\frac{A^2_{2S_{\nicefrac{1}{2}}}}{2}+\left(\frac{A_{2S_{\nicefrac{1}{2}}}}{4}-\frac{1}{2}\left(g_S\mu_B+g_I\mu_k\right)B_0\right)^2}}\nonumber.
\end{eqnarray}
\end{widetext}
Subsequently, follows the result for the $2P_{\nicefrac{1}{2}}$ set
\begin{widetext}
\begin{eqnarray}
    &&\ket{e_1}=\Ket{\nicefrac{3}{2},\nicefrac{3}{2},\nicefrac{1}{2},1}\nonumber\\
    &&E_{e_1}=-\Delta E_{Lamb}\frac{A_{2P_{\nicefrac{1}{2}}}}{2}+\left(\frac{g_{J_{2P_{\nicefrac{1}{2}}}}\mu_B}{2}-g_I\mu_k\right)B_0\nonumber\\
    &&\ket{e_2}=\frac{1}{\sqrt{1+\tilde{\gamma}_1^2(B)}}\left(\tilde{\gamma}_1(B)\ket{\nicefrac{1}{2},\nicefrac{1}{2},\nicefrac{1}{2},1}+\ket{\nicefrac{3}{2},\nicefrac{1}{2},\nicefrac{1}{2},1}\right)\nonumber\\
    &&E_{e_2}=-\Delta E_{Lamb}-\frac{A_{2P_{\nicefrac{1}{2}}}}{4}-\frac{g_I\mu_kB_0}{2}+\sqrt{\frac{A^2_{2P_{\nicefrac{1}{2}}}}{2}+\left(\frac{A_{2P_{\nicefrac{1}{2}}}}{4}+\frac{1}{2}\left(g_{J_{2P_{\nicefrac{1}{2}}}}\mu_B+g_I\mu_k\right)B_0\right)^2}\nonumber\\
    &&\ket{e_3}=\frac{1}{\sqrt{1+\tilde{\Gamma}_1^2(B)}}\left(\tilde{\Gamma}_1(B)\ket{\nicefrac{1}{2},-\nicefrac{1}{2},\nicefrac{1}{2},1}+\ket{\nicefrac{3}{2},-\nicefrac{1}{2},\nicefrac{1}{2},1}\right)\nonumber\\
    &&E_{e_3}=-\Delta E_{Lamb}-\frac{A_{2P_{\nicefrac{1}{2}}}}{4}+\frac{g_I\mu_kB_0}{2}+\sqrt{\frac{A^2_{2P_{\nicefrac{1}{2}}}}{2}+\left(\frac{A_{2P_{\nicefrac{1}{2}}}}{4}-\frac{1}{2}\left(g_{J_{2P_{\nicefrac{1}{2}}}}\mu_B+g_I\mu_k\right)B_0\right)^2}\nonumber\\
    &&\ket{f_4}=\Ket{\nicefrac{3}{2},-\nicefrac{3}{2},\nicefrac{1}{2},1}\nonumber\\
    &&E_{f_4}=-\Delta E_{Lamb}\frac{A_{2P_{\nicefrac{1}{2}}}}{2}-\left(\frac{g_{J_{2P_{\nicefrac{1}{2}}}}\mu_B}{2}-g_I\mu_k\right)B_0\nonumber\\
    &&\ket{f_5}=\frac{1}{\sqrt{1+\tilde{\Gamma}_2^2(B)}}\left(\ket{\nicefrac{1}{2},-\nicefrac{1}{2},\nicefrac{1}{2},1}+\tilde{\Gamma}_2(B)\ket{\nicefrac{3}{2},-\nicefrac{1}{2},\nicefrac{1}{2},1}\right)\nonumber\\
    &&E_{f_5}=-\Delta E_{Lamb}-\frac{A_{2P_{\nicefrac{1}{2}}}}{4}+\frac{g_I\mu_kB_0}{2}-\sqrt{\frac{A^2_{2P_{\nicefrac{1}{2}}}}{2}+\left(\frac{A_{2P_{\nicefrac{1}{2}}}}{4}-\frac{1}{2}\left(g_{J_{2P_{\nicefrac{1}{2}}}}\mu_B+g_I\mu_k\right)B_0\right)^2}\nonumber\\
    &&\ket{f_6}=\frac{1}{\sqrt{1+\tilde{\gamma}_2^2(B)}}\left(\ket{\nicefrac{1}{2},\nicefrac{1}{2},\nicefrac{1}{2},1}+\tilde{\gamma}_2(B)\ket{\nicefrac{3}{2},\nicefrac{1}{2},\nicefrac{1}{2},1}\right)\nonumber\\
    &&E_{f_6}=-\Delta E_{Lamb}-\frac{A_{2P_{\nicefrac{1}{2}}}}{4}-\frac{g_I\mu_kB_0}{2}-\sqrt{\frac{A^2_{2P_{\nicefrac{1}{2}}}}{2}+\left(\frac{A_{2P_{\nicefrac{1}{2}}}}{4}+\frac{1}{2}\left(g_{J_{2P_{\nicefrac{1}{2}}}}\mu_B+g_I\mu_k\right)B_0\right)^2}\nonumber\\
    &&\text{with}\quad \tilde{\gamma}_{\nicefrac{1}{2}}(B)=\mp\frac{\sqrt{2}}{3}\frac{\left(g_{J_{2P_{\nicefrac{1}{2}}}}\mu_B+g_I\mu_k\right)B_0}{\frac{3A_{2P_{\nicefrac{1}{2}}}}{4}+\frac{1}{6}\left(g_{J_{2P_{\nicefrac{1}{2}}}}\mu_B+g_I\mu_k\right)B_0+\sqrt{\frac{A^2_{2P_{\nicefrac{1}{2}}}}{2}+\left(\frac{A_{2P_{\nicefrac{1}{2}}}}{4}+\frac{1}{2}\left(g_{J_{2P_{\nicefrac{1}{2}}}}\mu_B+g_I\mu_k\right)B_0\right)^2}}\nonumber\\
    &&\text{with}\quad \tilde{\Gamma}_{\nicefrac{1}{2}}(B)=\mp\frac{\sqrt{2}}{3}\frac{\left(g_{J_{2P_{\nicefrac{1}{2}}}}\mu_B+g_I\mu_k\right)B_0}{\frac{3A_{2P_{\nicefrac{1}{2}}}}{4}-\frac{1}{6}\left(g_{J_{2P_{\nicefrac{1}{2}}}}\mu_B+g_I\mu_k\right)B_0+\sqrt{\frac{A^2_{2P_{\nicefrac{1}{2}}}}{2}+\left(\frac{A_{2P_{\nicefrac{1}{2}}}}{4}-\frac{1}{2}\left(g_{J_{2P_{\nicefrac{1}{2}}}}\mu_B+g_I\mu_k\right)B_0\right)^2}}\nonumber.
\end{eqnarray}
\end{widetext}
Finally, the result for the energetically higher positioned $2P_{\nicefrac{3}{2}}$ is given
\begin{widetext}
\begin{eqnarray}
    &&\ket{g_1}=\Ket{\nicefrac{5}{2},\nicefrac{5}{2},\nicefrac{3}{2},1}\nonumber\\
    &&E_{g_1}=FS+\frac{3A_{2P_{\nicefrac{3}{2}}}}{2}+\left(\frac{3g_{J_{2P_{\nicefrac{3}{2}}}}\mu_B}{2}-g_I\mu_k\right)B_0\nonumber\\
    &&\ket{g_2}=\frac{1}{\sqrt{1+\theta_1^2(B)}}\left(\theta_1(B)\ket{\nicefrac{3}{2},\nicefrac{3}{2},\nicefrac{3}{2},1}+\ket{\nicefrac{5}{2},\nicefrac{3}{2},\nicefrac{3}{2},1}\right)\nonumber\\
    &&E_{g_2}=FS+\frac{A_{2P_{\nicefrac{3}{2}}}}{4}+g_{J_{2P_{\nicefrac{3}{2}}}}\mu_BB_0-\frac{g_I\mu_kB_0}{2}+\sqrt{\frac{1}{4}\left(g_{J_{2P_{\nicefrac{3}{2}}}}\mu_B+g_I\mu_k\right)^2B_0^2-\frac{A_{2P_{\nicefrac{3}{2}}}}{4}\left(g_{J_{2P_{\nicefrac{3}{2}}}}\mu_B+g_I\mu_k\right)B_0+\frac{25}{16}A^2_{2P_{\nicefrac{3}{2}}}}\nonumber\\
    &&\ket{g_3}=\frac{1}{\sqrt{1+\alpha_1^2(B)+\beta_1^2(B)}}\left(\beta_1(B)\ket{\nicefrac{1}{2},\nicefrac{1}{2},\nicefrac{3}{2},1}+\alpha_1(B)\ket{\nicefrac{3}{2},\nicefrac{1}{2},\nicefrac{3}{2},1}+\ket{\nicefrac{5}{2},\nicefrac{1}{2},\nicefrac{3}{2},1}\right)\nonumber\\
    &&E_{g_3}=FS+\frac{1}{30}\left(-20A_{2P_{\nicefrac{3}{2}}}+15g_{J_{2P_{\nicefrac{3}{2}}}}\mu_BB_0+\frac{245A^2_{2P_{\nicefrac{3}{2}}}-60A_{2P_{\nicefrac{3}{2}}}\left(g_{J_{2P_{\nicefrac{3}{2}}}}\mu_B+g_I\mu_k\right)B_0+60\left(g_{J_{2P_{\nicefrac{3}{2}}}}\mu_B+g_I\mu_k\right)^2B_0^2}{C^{\nicefrac{1}{3}}}+5C^{\nicefrac{1}{3}}\right)\nonumber\\
    &&\ket{g_4}=\frac{1}{\sqrt{1+\tilde{\alpha}_1^2(B)+\tilde{\beta}_1^2(B)}}\left(\tilde{\beta}_1(B)\ket{\nicefrac{1}{2},-\nicefrac{1}{2},\nicefrac{3}{2},1}+\tilde{\alpha}_1(B)\ket{\nicefrac{3}{2},-\nicefrac{1}{2},\nicefrac{3}{2},1}+\ket{\nicefrac{5}{2},-\nicefrac{1}{2},\nicefrac{3}{2},1}\right)\nonumber\\
    &&E_{g_4}=FS+\frac{1}{30}\left(-5\left(4A_{2P_{\nicefrac{3}{2}}}+3g_{J_{2P_{\nicefrac{3}{2}}}}\mu_BB_0\right)+\frac{245A^2_{2P_{\nicefrac{3}{2}}}+60A_{2P_{\nicefrac{3}{2}}}\left(g_{J_{2P_{\nicefrac{3}{2}}}}\mu_B+g_I\mu_k\right)B_0+60\left(g_{J_{2P_{\nicefrac{3}{2}}}}\mu_B+g_I\mu_k\right)^2B_0^2}{\tilde{C}^{\nicefrac{1}{3}}}+5\tilde{C}^{\nicefrac{1}{3}}\right)\nonumber\\
    &&\ket{g_5}=\frac{1}{\sqrt{1+\phi_1^2(B)}}\left(\phi_1(B)\ket{\nicefrac{3}{2},-\nicefrac{3}{2},\nicefrac{3}{2},1}+\ket{\nicefrac{5}{2},-\nicefrac{3}{2},\nicefrac{3}{2},1}\right)\nonumber\\
    &&E_{g_5}=FS+\frac{A_{2P_{\nicefrac{3}{2}}}}{4}-g_{J_{2P_{\nicefrac{3}{2}}}}\mu_BB_0+\frac{g_I\mu_kB_0}{2}+\sqrt{\frac{1}{4}\left(g_{J_{2P_{\nicefrac{3}{2}}}}\mu_B+g_I\mu_k\right)^2B_0^2+\frac{A_{2P_{\nicefrac{3}{2}}}}{4}\left(g_{J_{2P_{\nicefrac{3}{2}}}}\mu_B+g_I\mu_k\right)B_0+\frac{25}{16}A^2_{2P_{\nicefrac{3}{2}}}}\nonumber\\
    &&\ket{g_6}=\ket{\nicefrac{5}{2},-\nicefrac{5}{2},\nicefrac{3}{2},1}\nonumber\\
    &&E_{g_6}=FS+\frac{3A_{2P_{\nicefrac{3}{2}}}}{2}-\left(\frac{3g_{J_{2P_{\nicefrac{3}{2}}}}\mu_B}{2}-g_I\mu_k\right)B_0\nonumber\\
    &&\ket{h_7}=\frac{1}{\sqrt{1+\theta_2^2(B)}}\left(\ket{\nicefrac{3}{2},\nicefrac{3}{2},\nicefrac{3}{2},1}+\theta_2(B)\ket{\nicefrac{5}{2},\nicefrac{3}{2},\nicefrac{3}{2},1}\right)\nonumber\\
    &&E_{h_7}=FS+\frac{A_{2P_{\nicefrac{3}{2}}}}{4}+g_{J_{2P_{\nicefrac{3}{2}}}}\mu_BB_0-\frac{g_I\mu_kB_0}{2}-\sqrt{\frac{1}{4}\left(g_{J_{2P_{\nicefrac{3}{2}}}}\mu_B+g_I\mu_k\right)^2B_0^2-\frac{A_{2P_{\nicefrac{3}{2}}}}{4}\left(g_{J_{2P_{\nicefrac{3}{2}}}}\mu_B+g_I\mu_k\right)B_0+\frac{25}{16}A^2_{2P_{\nicefrac{3}{2}}}}\nonumber\\
    &&\ket{h_8}=\frac{1}{\sqrt{1+\alpha_2^2(B)+\beta_2^2(B)}}\left(\beta_2(B)\ket{\nicefrac{1}{2},\nicefrac{1}{2},\nicefrac{3}{2},1}+\ket{\nicefrac{3}{2},\nicefrac{1}{2},\nicefrac{3}{2},1}+\alpha_2(B)\ket{\nicefrac{5}{2},\nicefrac{1}{2},\nicefrac{3}{2},1}\right)\nonumber\\
    &&E_{h_8}=FS+\frac{1}{12}\left(-8A_{2P_{\nicefrac{3}{2}}}+6g_{J_{2P_{\nicefrac{3}{2}}}}\mu_BB_0+\right.\nonumber\\&&\left.+\frac{49i\left(i+\sqrt{3}\right)A^2_{2P_{\nicefrac{3}{2}}}+12\left(1-\sqrt{3}i\right)A_{2P_{\nicefrac{3}{2}}}\left(g_{J_{2P_{\nicefrac{3}{2}}}}\mu_B+g_I\mu_k\right)B_0+12i\left(i+\sqrt{3}\right)\left(g_{J_{2P_{\nicefrac{3}{2}}}}\mu_B+g_I\mu_k\right)^2B_0^2}{C^{\nicefrac{1}{3}}}-i\left(-i+\sqrt{3}\right)C^{\nicefrac{1}{3}}\right)\nonumber\\
    &&\ket{h_9}=\frac{1}{\sqrt{1+\tilde{\alpha}_2^2(B)+\tilde{\beta}_2^2(B)}}\left(\tilde{\beta}_2(B)\ket{\nicefrac{1}{2},-\nicefrac{1}{2},\nicefrac{3}{2},1}+\ket{\nicefrac{3}{2},-\nicefrac{1}{2},\nicefrac{3}{2},1}+\tilde{\alpha}_2(B)\ket{\nicefrac{5}{2},-\nicefrac{1}{2},\nicefrac{3}{2},1}\right)\nonumber\\
    &&E_{h_9}=FS+\frac{1}{30}\left(-5\left(4A_{2P_{\nicefrac{3}{2}}}+3g_{J_{2P_{\nicefrac{3}{2}}}}\mu_BB_0\right)+\right.\nonumber\\&&\left.+\frac{5i\left(i+\sqrt{3}\right)\left(49A^2_{2P_{\nicefrac{3}{2}}}+12A_{2P_{\nicefrac{3}{2}}}\left(g_{J_{2P_{\nicefrac{3}{2}}}}\mu_B+g_I\mu_k\right)B_0+12\left(g_{J_{2P_{\nicefrac{3}{2}}}}\mu_B+g_I\mu_k\right)^2B_0^2\right)}{2\tilde{C}^{\nicefrac{1}{3}}}-\frac{5}{2}i\left(-i+\sqrt{3}\right)\tilde{C}^{\nicefrac{1}{3}}\right)\nonumber
\end{eqnarray}
\end{widetext}
\begin{widetext}
\begin{eqnarray}
    &&\ket{h_{10}}=\frac{1}{\sqrt{1+\phi_2^2(B)}}\left(\ket{\nicefrac{3}{2},-\nicefrac{3}{2},\nicefrac{3}{2},1}+\phi_2(B)\ket{\nicefrac{5}{2},-\nicefrac{3}{2},\nicefrac{3}{2},1}\right)\nonumber\\
    &&E_{h_{10}}=FS+\frac{A_{2P_{\nicefrac{3}{2}}}}{4}-g_{J_{2P_{\nicefrac{3}{2}}}}\mu_BB_0+\frac{g_I\mu_kB_0}{2}-\sqrt{\frac{1}{4}\left(g_{J_{2P_{\nicefrac{3}{2}}}}\mu_B+g_I\mu_k\right)^2B_0^2+\frac{A_{2P_{\nicefrac{3}{2}}}}{4}\left(g_{J_{2P_{\nicefrac{3}{2}}}}\mu_B+g_I\mu_k\right)B_0+\frac{25}{16}A^2_{2P_{\nicefrac{3}{2}}}}\nonumber\\
    &&\ket{k_{11}}=\frac{1}{\sqrt{1+\alpha_3^2(B)+\beta_3^2(B)}}\left(\ket{\nicefrac{1}{2},\nicefrac{1}{2},\nicefrac{3}{2},1}+\alpha_3(B)\ket{\nicefrac{3}{2},\nicefrac{1}{2},\nicefrac{3}{2},1}+\beta_3(B)\ket{\nicefrac{5}{2},\nicefrac{1}{2},\nicefrac{3}{2},1}\right)\nonumber\\
    &&E_{k_{11}}=FS+\frac{1}{12}\left(-8A_{2P_{\nicefrac{3}{2}}}+6g_{J_{2P_{\nicefrac{3}{2}}}}\mu_BB_0-\right.\nonumber\\&&\left.-\frac{i\left(-i+\sqrt{3}\right)\left(49A^2_{2P_{\nicefrac{3}{2}}}-12A_{2P_{\nicefrac{3}{2}}}\left(g_{J_{2P_{\nicefrac{3}{2}}}}\mu_B+g_I\mu_k\right)B_0+12\left(g_{J_{2P_{\nicefrac{3}{2}}}}\mu_B+g_I\mu_k\right)^2B_0^2\right)}{C^{\nicefrac{1}{3}}}+i\left(i+\sqrt{3}\right)C^{\nicefrac{1}{3}}\right)\nonumber\\
    &&\ket{k_{12}}=\frac{1}{\sqrt{1+\tilde{\alpha}_3^2(B)+\tilde{\beta}_3^2(B)}}\left(\ket{\nicefrac{1}{2},-\nicefrac{1}{2},\nicefrac{3}{2},1}+\tilde{\alpha}_3(B)\ket{\nicefrac{3}{2},-\nicefrac{1}{2},\nicefrac{3}{2},1}+\tilde{\beta}_3(B)\ket{\nicefrac{5}{2},-\nicefrac{1}{2},\nicefrac{3}{2},1}\right)\nonumber\\
    &&E_{k_{12}}=FS+\frac{1}{30}\left(-5\left(4A_{2P_{\nicefrac{3}{2}}}+3g_{J_{2P_{\nicefrac{3}{2}}}}\mu_BB_0\right)-\right.\nonumber\\&&\left.-\frac{5i\left(-i+\sqrt{3}\right)\left(49A^2_{2P_{\nicefrac{3}{2}}}+12A_{2P_{\nicefrac{3}{2}}}\left(g_{J_{2P_{\nicefrac{3}{2}}}}\mu_B+g_I\mu_k\right)B_0+12\left(g_{J_{2P_{\nicefrac{3}{2}}}}\mu_B+g_I\mu_k\right)^2B_0^2\right)}{2\tilde{C}^{\nicefrac{1}{3}}}+\frac{5}{2}i\left(i+\sqrt{3}\right)\tilde{C}^{\nicefrac{1}{3}}\right)\nonumber\\
    &&\text{with}\quad \theta_{\nicefrac{1}{2}}(B)=\mp\frac{\sqrt{6}}{5}\frac{\left(g_{J_{2P_{\nicefrac{3}{2}}}}\mu_B+g_I\mu_k\right)B_0}{\frac{5A_{2P_{\nicefrac{3}{2}}}}{4}-\frac{1}{10}\left(g_{J_{2P_{\nicefrac{3}{2}}}}\mu_B+g_I\mu_k\right)B_0+\sqrt{\frac{1}{4}\left(g_{J_{2P_{\nicefrac{3}{2}}}}\mu_B+g_I\mu_k\right)^2B_0^2-\frac{A_{2P_{\nicefrac{3}{2}}}}{4}\left(g_{J_{2P_{\nicefrac{3}{2}}}}\mu_B+g_I\mu_k\right)B_0+\frac{25}{16}A^2_{2P_{\nicefrac{3}{2}}}}}\nonumber\\
    &&\text{with}\quad \phi_{\nicefrac{1}{2}}(B)=\mp\frac{\sqrt{6}}{5}\frac{\left(g_{J_{2P_{\nicefrac{3}{2}}}}\mu_B+g_I\mu_k\right)B_0}{\frac{5A_{2P_{\nicefrac{3}{2}}}}{4}+\frac{1}{10}\left(g_{J_{2P_{\nicefrac{3}{2}}}}\mu_B+g_I\mu_k\right)B_0+\sqrt{\frac{1}{4}\left(g_{J_{2P_{\nicefrac{3}{2}}}}\mu_B+g_I\mu_k\right)^2B_0^2+\frac{A_{2P_{\nicefrac{3}{2}}}}{4}\left(g_{J_{2P_{\nicefrac{3}{2}}}}\mu_B+g_I\mu_k\right)B_0+\frac{25}{16}A^2_{2P_{\nicefrac{3}{2}}}}}\nonumber\\
    &&\text{with}\quad C=143A^3_{2P_{\nicefrac{3}{2}}}+18A^2_{2P_{\nicefrac{3}{2}}}\left(g_{J_{2P_{\nicefrac{3}{2}}}}\mu_B+g_I\mu_k\right)B_0-72A_{2P_{\nicefrac{3}{2}}}\left(g_{J_{2P_{\nicefrac{3}{2}}}}\mu_B+g_I\mu_k\right)^2B_0^2+\nonumber\\&&\qquad+\sqrt{A^2_{2P_{\nicefrac{3}{2}}}\left(143A^2_{2P_{\nicefrac{3}{2}}}+18A_{2P_{\nicefrac{3}{2}}}\left(g_{J_{2P_{\nicefrac{3}{2}}}}\mu_B+g_I\mu_k\right)B_0-72\left(g_{J_{2P_{\nicefrac{3}{2}}}}\mu_B+g_I\mu_k\right)^2B_0^2\right)^2-}\nonumber\\&&\qquad\overline{\rule{0pt}{2.5ex}-\left(49A^2_{2P_{\nicefrac{3}{2}}}-12A_{2P_{\nicefrac{3}{2}}}\left(g_{J_{2P_{\nicefrac{3}{2}}}}\mu_B+g_I\mu_k\right)B_0+12\left(g_{J_{2P_{\nicefrac{3}{2}}}}\mu_B+g_I\mu_k\right)^2B_0^2\right)^3}\nonumber\\
    &&\text{with}\quad \tilde{C}=143A^3_{2P_{\nicefrac{3}{2}}}-18A^2_{2P_{\nicefrac{3}{2}}}\left(g_{J_{2P_{\nicefrac{3}{2}}}}\mu_B+g_I\mu_k\right)B_0-72A_{2P_{\nicefrac{3}{2}}}\left(g_{J_{2P_{\nicefrac{3}{2}}}}\mu_B+g_I\mu_k\right)^2B_0^2+\nonumber\\&&\qquad+\sqrt{A^2_{2P_{\nicefrac{3}{2}}}\left(-143A^2_{2P_{\nicefrac{3}{2}}}+18A_{2P_{\nicefrac{3}{2}}}\left(g_{J_{2P_{\nicefrac{3}{2}}}}\mu_B+g_I\mu_k\right)B_0+72\left(g_{J_{2P_{\nicefrac{3}{2}}}}\mu_B+g_I\mu_k\right)^2B_0^2\right)^2-}\nonumber\\&&\qquad\overline{\rule{0pt}{2.5ex}-\left(49A^2_{2P_{\nicefrac{3}{2}}}+12A_{2P_{\nicefrac{3}{2}}}\left(g_{J_{2P_{\nicefrac{3}{2}}}}\mu_B+g_I\mu_k\right)B_0+12\left(g_{J_{2P_{\nicefrac{3}{2}}}}\mu_B+g_I\mu_k\right)^2B_0^2\right)^3}\nonumber\\
    &&\text{with}\quad \nicefrac{\alpha_1}{\tilde{\alpha}}_1(B)=-\frac{3}{5}\frac{\left(g_{J_{2P_{\nicefrac{3}{2}}}}\mu_B+g_I\mu_k\right)B_0}{-A_{2P_{\nicefrac{3}{2}}}+\left(\pm\frac{11}{30}g_{J_{2P_{\nicefrac{3}{2}}}}\mu_B\mp\frac{2}{15}g_I\mu_k\right)B_0-\left(E_{\nicefrac{g_3}{g_4}}-FS\right)-\frac{5}{9}\frac{\left(g_{J_{2P_{\nicefrac{3}{2}}}}\mu_B+g_I\mu_k\right)^2B_0^2}{-\frac{5A_{2P_{\nicefrac{3}{2}}}}{2}\pm\left(\frac{5}{6}g_{J_{2P_{\nicefrac{3}{2}}}}\mu_B+\frac{1}{3}g_I\mu_k\right)B_0-\left(E_{\nicefrac{g_3}{g_4}}-FS\right)}}\nonumber\\
    &&\text{with}\quad \nicefrac{\alpha_2}{\tilde{\alpha}}_2(B)=-\frac{3}{5}\frac{\left(g_{J_{2P_{\nicefrac{3}{2}}}}\mu_B+g_I\mu_k\right)B_0}{\frac{3A_{2P_{\nicefrac{3}{2}}}}{2}\pm\frac{1}{5}\left(\frac{3}{2}g_{J_{2P_{\nicefrac{3}{2}}}}\mu_B-g_I\mu_k\right)B_0-\left(E_{\nicefrac{h_8}{h_9}}-FS\right)}\nonumber.
\end{eqnarray}
\end{widetext}
\begin{widetext}
\begin{eqnarray}
    &&\text{with}\quad \nicefrac{\alpha_3}{\tilde{\alpha}}_3(B)=-\frac{\sqrt{5}}{3}\frac{\left(g_{J_{2P_{\nicefrac{3}{2}}}}\mu_B+g_I\mu_k\right)B_0}{-A_{2P_{\nicefrac{3}{2}}}+\left(\pm\frac{11}{30}g_{J_{2P_{\nicefrac{3}{2}}}}\mu_B\mp\frac{2}{15}g_I\mu_k\right)B_0-\left(E_{\nicefrac{k_{11}}{k_{12}}}-FS\right)-\frac{9}{25}\frac{\left(g_{J_{2P_{\nicefrac{3}{2}}}}\mu_B+g_I\mu_k\right)^2B_0^2}{\frac{3A_{2P_{\nicefrac{3}{2}}}}{2}\pm\frac{1}{5}\left(\frac{3}{2}g_{J_{2P_{\nicefrac{3}{2}}}}\mu_B-g_I\mu_k\right)B_0-\left(E_{\nicefrac{k_{11}}{k_{12}}}-FS\right)}}\nonumber\\
    &&\text{with}\quad \nicefrac{\beta_1}{\tilde{\beta}}_1(B)=\frac{1}{\sqrt{5}}\frac{\left(g_{J_{2P_{\nicefrac{3}{2}}}}\mu_B+g_I\mu_k\right)^2B_0^2}{\left(-A_{2P_{\nicefrac{3}{2}}}+\left(\pm\frac{11}{30}g_{J_{2P_{\nicefrac{3}{2}}}}\mu_B\mp\frac{2}{15}g_I\mu_k\right)B_0-\left(E_{\nicefrac{g_{3}}{g_{4}}}-FS\right)\right)\cdot}\nonumber\\&&\frac{1}{\cdot\left(-\frac{5A_{2P_{\nicefrac{3}{2}}}}{2}+\left(\pm\frac{5}{6}g_{J_{2P_{\nicefrac{3}{2}}}}\mu_B\mp\frac{1}{3}g_I\mu_k\right)B_0-\left(E_{\nicefrac{g_{3}}{g_{4}}}-FS\right)\right)-\frac{5}{9}\left(g_{J_{2P_{\nicefrac{3}{2}}}}\mu_B+g_I\mu_k\right)^2B_0^2}\nonumber\\
    &&\text{with}\quad \nicefrac{\beta_2}{\tilde{\beta}}_2(B)=-\frac{\sqrt{5}}{3}\frac{\left(g_{J_{2P_{\nicefrac{3}{2}}}}\mu_B+g_I\mu_k\right)B_0}{-\frac{5A_{2P_{\nicefrac{3}{2}}}}{2}\pm\left(\frac{5}{6}g_{J_{2P_{\nicefrac{3}{2}}}}\mu_B+\frac{1}{3}g_I\mu_k\right)B_0-\left(E_{\nicefrac{h_{8}}{h_{9}}}-FS\right)}\nonumber\\
    &&\text{with}\quad \nicefrac{\beta_3}{\tilde{\beta}}_3(B)=\frac{1}{\sqrt{5}}\frac{\left(g_{J_{2P_{\nicefrac{3}{2}}}}\mu_B+g_I\mu_k\right)^2B_0^2}{\left(-A_{2P_{\nicefrac{3}{2}}}+\left(\pm\frac{11}{30}g_{J_{2P_{\nicefrac{3}{2}}}}\mu_B\mp\frac{2}{15}g_I\mu_k\right)B_0-\left(E_{\nicefrac{k_{11}}{k_{12}}}-FS\right)\right)\cdot}\nonumber\\&&\frac{1}{\cdot\left(\frac{3A_{2P_{\nicefrac{3}{2}}}}{2}+\frac{1}{5}\left(\pm\frac{3}{2}g_{J_{2P_{\nicefrac{3}{2}}}}\mu_B\mp g_I\mu_k\right)B_0-\left(E_{\nicefrac{k_{11}}{k_{12}}}-FS\right)\right)-\frac{9}{25}\left(g_{J_{2P_{\nicefrac{3}{2}}}}\mu_B+g_I\mu_k\right)^2B_0^2}\nonumber.
\end{eqnarray}
\end{widetext}
\section{Landé factor}\label{g_j}
The orbital angular momentum $\vec{L}$ and the electron spin $\vec{S}$ are combined to the total angular momentum $\vec{J}=\vec{L}\otimes\mathds{1}+\mathds{1}\otimes\vec{S}$ of the electron. Since those angular momenta create magnetic moments which are proportional to their anomalous g-factors $g_{\nicefrac{l}{s}}$ one needs to combine these to a new g-factor for $\vec{J}$ called the Landé factor and is given by 
\begin{widetext}
\begin{eqnarray}
    g_j=\frac{1}{2}\frac{g_s\left[J\left(J+1\right)+S\left(S+1\right)-L\left(L+1\right)\right]+g_l\left[J\left(J+1\right)+L\left(L+1\right)-S\left(S+1\right)\right]}{J\left(J+1\right)}.
\end{eqnarray}
\end{widetext}
The g-factors then equal to $g_s\approx2.002$~\cite{g_s_neu} and $g_l=1$~\cite{g_s_alt} which leads to the different Landé factors of $g_{\nicefrac{1}{2}}=\frac{1}{3}\left(4-g_s\right)$ and $g_{\nicefrac{3}{2}}=\frac{1}{3}\left(2+g_s\right)$ for the different sets of states.
\section{Hyperfine constant}\label{A}
For hyperfine transitions between states from the same set of states the Hamiltonian describing the interaction can be reduced to
\begin{equation}
    H_{Hyp}=A\frac{\vec{I}\cdot\vec{J}}{\hbar^2}.
\end{equation}
\newline
$A$ is then the hyperfine splitting constant, which can be calculated theoretically to first order~\cite{Bethe} for $L=0$ by
\begin{equation}
    A=\frac{e^2g_I\hbar^2}{3\varepsilon_0c^2m_em_p}\frac{1}{4\pi}|R_{n,0}(r=0)|^2,
\end{equation}
and for $L\neq0$ by
\begin{equation}
  A=\frac{e^2\hbar^2g_I}{2m_em_p4\pi\varepsilon_0c^2j(j+1)a_0^3n^3\left(l+\nicefrac{1}{2}\right)}.  
\end{equation}
\nocite{*}
\newpage
\bibliography{apssamp}

\end{document}